\documentclass[%
reprint,
superscriptaddress,
nofootinbib,       %
 amsmath,amssymb,
 aps, prb,
]{revtex4-2}

\newcommand*\diff{\mathop{}\!\mathrm{d}}

\usepackage[T1]{fontenc}
\usepackage{siunitx}
\usepackage{graphicx}
\usepackage{layouts}
\usepackage{booktabs}
\usepackage{multirow}
\usepackage{hyperref}
\newcommand{\dbend}{\"{u}}
\newcommand{\be}{\begin{equation}}
\newcommand{\ee}{\end{equation}}
\usepackage{xcolor}
\usepackage{gensymb}

\definecolor{seabornBlue}{RGB}{76,114,176}
\definecolor{seabornGreen}{RGB}{85,168,104}
\definecolor{seabornRed}{RGB}{196,78,82}

\definecolor{orangePumpkin}{RGB}{211,84,0}
\definecolor{orangeCarrot}{RGB}{230,126,34}
\definecolor{blueBelizeHole}{RGB}{41,128,185}
\definecolor{redAlizarin}{RGB}{231,76,60}
\definecolor{redNasturcianFlower}{RGB}{232,65,24}

\begin{document}
\vspace{-0.5cm}

\title{
Increasing valley splitting in Si/SiGe by practically achievable heterostructure profiles}

\author{Lukas Cvitkovich}
\email[]{l.cvitkovich@gmail.com}
\affiliation{Fakultät für Physik, Universität Regensburg, 93040 Regensburg, Germany}
\affiliation{ResQ Resonant Quantum Systems FlexCo, 7061 Trausdorf an der Wulka, Austria}

\author{Peter Stano}
\email[]{peter.stano@riken.jp}
\affiliation{RIKEN, Center for Quantum Computing (RQC), Wako-shi, Saitama 351-0198, Japan}
\affiliation{Institute of Physics, Slovak Academy of Sciences, Bratislava, Slovakia}

\author{Dominique Bougeard}
\affiliation{Fakultät für Physik, Universität Regensburg, 93040 Regensburg, Germany}

\author{Yann-Michel Niquet}
\affiliation{Univ. Grenoble Alpes, CEA, IRIG-MEM-L\_Sim, F-38000, Grenoble, France}

\author{Daniel Loss}
\affiliation{RIKEN, Center for Quantum Computing (RQC), Wako-shi, Saitama 351-0198, Japan}
\affiliation{Physics Department, King Fahd University of Petroleum and Minerals, 31261, Dhahran, Saudi Arabia}
\affiliation{Center for Advanced Quantum Computing, KFUPM, Dhahran, Saudi Arabia}
\affiliation{RDIA Chair in Quantum Computing}
\affiliation{Department of Physics, University of Basel, Klingelbergstrasse 82, CH-4056 Basel, Switzerland}

\begin{abstract}
Silicon spin qubits are marred by the valley degeneracy of the conduction band. In a nanodevice, the degeneracy is lifted by interfaces and alloy disorder, but the resulting valley splitting is small, of order 100 $\mu$eV in Si/SiGe quantum wells.  
Substantial efforts were invested both in theory and experiments to overcome the valley issue. Unfortunately, the existing recipes either rely on atomistic details of the interface that are beyond experimental control, or demand heterostructure profiles beyond current state-of-the-art heterostructure epitaxy.
We revisit the valley splitting induced by non-trivial Ge profiles and advocate a novel view of the intervalley coupling as backscattering on point-like impurities realized by crystal planes containing Ge atoms. This perspective reveals that enhancing the backscattering amplitude, which sets the valley splitting, requires constructive interference of multiple scatterers. %
We arrive at a remarkable prediction that the Ge content along the heterostructure growth direction does not have to have any specific periodicity, including the practically unreachable $2\pi/(2k_0)$ period, to significantly increase the valley splitting. 
Numerical evidence from tight-binding simulations and intuitive physical interpretations corroborate this statement. We devise profiles that seem within the capabilities of current MBE growth techniques and boost the valley splitting beyond the 1\,meV scale.

\end{abstract}

\maketitle

\section{Importance of the valley splitting}
Electron spin qubits in silicon are promising for quantum computing~\cite{Burkard1999, Zwanenburg2013}. They have long coherence times due to weak spin-orbit coupling~\cite{prada_spinorbit_2011, hsueh_engineering_2024} and the possibility to suppress nuclear spin noise by isotopic purification~\cite{itoh_isotope_2014, Cvitkovich_hf, Wild_purified_2021}. While scaling up is a challenge for every platform, silicon benefits from an established technological infrastructure~\cite{Huckemann_industrial_2025, Koch2025, George_12qubits_300mm_2025}.

A key drawback of silicon is its conduction band degeneracy. While the bulk six-fold degeneracy is partially lifted by strain\footnote{See Tab.~I in Ref.~\cite{balslev_influence_1966}; Fig.~1 in Ref.~\cite{zeller_electric_1986}; Fig.~7 in Ref.~\cite{abstreiter_silicongermanium_1989}.} and two-dimensional electron gas (2DEG) confinement\footnote{See around Eq. (35) in \cite{sun_physics_2007}.}, the remaining two-fold degeneracy is a roadblock for spin qubits.\footnote{Ref.~\cite{culcer_realizing_2009}: ``The biggest obstacle to spin QC in Si is valley degeneracy''. Ref.~\cite{friesen_theory_2010}: ``small gaps of any type are anathema for qubit operations.''} The associated valley degree of freedom introduces a host of problems \cite{koiller_exchange_2001, culcer_realizing_2009, friesen_theory_2010, Zwanenburg2013, losert_strategies_2024}, such as Pauli spin blockade lifting, qubit leakage, spin relaxation, spin-spin exchange volatility, spin dephasing due to the valley dependence of the g-factor, or fidelity loss upon electron shuttling.

In a nanodevice, the remaining two-fold degeneracy is lifted since, with the quantum well interfaces, electric fields from gates, and alloy disorder, none of the original symmetries of the crystal is preserved, and all degeneracies will be, strictly speaking, split. However, the resulting energy splitting of the two valley states, the valley splitting, is low in practice. In Si/SiGe,
sensitivity to barrier details\footnote{``The breaking of the two-fold valley degeneracy is very sensitive to atomic-scale details of the interface'' \cite{Zwanenburg2013}}, disorder, and atomistic fluctuations~\cite{Wuetz2022}, lead to a large spread of observed values, varying from 20 to 300\,$\mu$eV \cite{Shi2011, Kawakami2014, Mi2017, Oh2021, Hollmann_largeVS_2020}.
Though the problem is less severe in MOS devices, for which somewhat larger values are quoted,\footnote{Ref.\cite{yang_spin-valley_2013} demonstrates electrically-tunable valley splitting within the range of 300--800 $\mu$eV; Ref.~\cite{veldhorst_addressable_2014} within 100--500 $\mu$eV.} Si/SiGe stands out as an exceptional 2DEG environment, believed to have lowest noise and defect levels, offering qubits with state-of-the-art coherence \cite{yoneda_quantum-dot_2017, Struck2020,mills_two-qubit_2022,weinstein_universal_2023} and fidelity \cite{takeda_resonantly_2020,blumoff_fast_2022,philips_universal_2022,mills_high-fidelity_2022}. Resolving the valley issue in Si/SiGe on the material level would be a major breakthrough for semiconducting spin qubits.

The early numerical investigations\footnote{See the Supplemental Material of Ref.~\cite{Zwanenburg2013} and the references therein.} and measurements \cite{goswami_controllable_2007} suggested that the valley splitting could be pushed to a meV scale in narrow quantum wells. However, these predictions, relying on atomistically sharp interfaces\footnote{We note here Ref.~\cite{zhang_genetic_2013}, which suggested obtaining sharp interfaces using Bragg-mirror barriers implemented by Si-Ge superlattices.}, were found unrealistic. The finite width and imperfections of the interface severely diminish the resulting valley splitting.\footnote{That even a slight misorientation of the interface ``enormously'' reduces the valley splitting has been suggested by Ando in Ref.~\cite{Ando_valley_1979} as an explanation of low valley splitting values observed in Hall-bar devices.} In current Si/SiGe structures with real-world imperfections, the valley splitting is probably dominated by the alloy disorder (that is, replacing some of the atoms of Si by Ge) inside the quantum well or near its interfaces, and is one or two orders of magnitude smaller than the early optimistic estimates.

\section{The physics of the valley splitting}

\newcommand{\PsiEnv}{\psi_\mathrm{env}}
\newcommand{\PsiFull}{\psi}
\newcommand{\PsiScattering}{\psi^\mathrm{sc}}
\newcommand{\PsiVariational}{\psi^\mathrm{var}}

\newcommand{\QWPot}{U_\mathrm{qw}}

\newcommand{\VS}{E_\mathrm{VS}}

\newcommand{\VGe}{V_\mathrm{Ge}}

\newcommand{\MEFFsymbol}{S}
\newcommand{\MEFFname}{structure factor}

Since the Si/SiGe interface alone does not give a large enough splitting,\footnote{``Theoretical predictions for the valley splitting of flat interfaces are generally on the order of 0.1--0.3 meV'' \cite{Zwanenburg2013}.} several proposals considered creating sharp features inside the Si well \cite{McJunkin2021, Feng2022, Wuetz2022, McJunkin2022, losert_practical_2023, sarkar_micromagnet-free_2025, thayil_optimization_2025} to provide the momentum required for intervalley scattering. The main idea can be understood from a simple model that we discuss next.

\subsection{The ``$2k_0$ theory'' \cite{losert_practical_2023}}

Let us consider a two-dimensional electron gas (2DEG) in a strained Si/SiGe heterostructure grown along the [001] crystal axis called the $z$ direction. We first take into account only the $z$ coordinate as the degree of freedom and add lateral coordinates later. There are two conduction-band minima, at crystal momenta $\pm k_0 \mathbf{z}$, which remain from the sixfold degeneracy of the bulk after imposing strain and quantum well confinement  $\QWPot$. Here we take\footnote{We choose 0.84 as an often quoted value from the interval 0.82--0.85 within which the minimum is located with high certainty. 
} $k_0 = 0.84 \times (2\pi / a_\mathrm{Si})$ (in further, we mostly omit the units $(2\pi/a_\mathrm{Si}$) when stating numerical values for $k_0$), with the bulk Si lattice parameter $a_\mathrm{Si}=0.543\,\mathrm{nm}$ \cite{Reeber1996}. Let us first take the two minima isolated.  At the level of the effective mass theory, they are identical and correspond to the same envelope wave-function ground state $\PsiEnv(z)$. Reinstating the Bloch parts $u_\pm(z)$, which have the translational symmetry of the crystal lattice, the full wave functions of the doubly-degenerate ground state are\footnote{We use the notation of Ref.~\cite{losert_practical_2023}.}
\be
\PsiFull_\pm (z) = \PsiEnv(z) e^{\pm i k_0 z} u_\pm(z).
\ee
Considering now the two minima together, they will be coupled by the off-diagonal matrix element of the effective-mass Hamiltonian \cite{Ando_valley_1979,Feng2022, Wuetz2022, losert_practical_2023} 
\be
\label{eq:Delta}
\Delta = \langle \PsiFull_{+} | \QWPot | \PsiFull_{-} \rangle.
\ee
In the lowest-order perturbation theory \cite{Ando_valley_1979, friesen_valley_2007}, the valley splitting is $\VS=2|\Delta|$. Extensive tight-binding calculations confirm that Eq.~\eqref{eq:Delta} provides a reliable estimate of the valley splitting \cite{losert_practical_2023, Salamone_kp_2026}.

We rewrite the above expression as the valley splitting being proportional to an integral
\be
\label{eq:VSisFT}
\VS \propto \left|  \int  \diff z |\PsiEnv (z)|^2 \QWPot(z) e^{-i 2k_0 z} \right|.
\ee
We have sacrificed the overall prefactor in favor of dropping the Bloch amplitudes $u_\pm$\footnote{Their overlap matrix elements have been given in Ref.~\cite{saraiva_intervalley_2011}. } and a freedom of a scale parameter in $\QWPot$ that we will introduce below. In any case, one arrives at a crucial observation. With the envelope function $\PsiEnv$ being smooth on atomic distances, the integral in Eq.~\eqref{eq:VSisFT} is dominated by sharp features of $\QWPot$. In other words, the valley splitting is simply given by the $2k_0$ Fourier component of the heterostructure potential. This is called the ``$2k_0$ theory'' \cite{losert_practical_2023}.

Equation \eqref{eq:VSisFT} explains the ideas in the proposals mentioned above and reviewed recently in Ref.~\cite{losert_practical_2023}. For example, Ref.~\cite{McJunkin2021} considers a single spike in $\QWPot$ placed at the maximum of $|\PsiEnv (z)|^2$, while the ``wiggle well'' suggestion of Ref.~\cite{McJunkin2022, Feng2022} entails sinusoidal oscillations at the matching wave vector $2k_0$,
 Again, putting $|\PsiEnv (z)|^2$ aside, such oscillatory $\QWPot$ is optimal in a well-defined sense, as a function with all Fourier components matching the required intervalley momentum difference $2k_0$ and no Fourier components elsewhere,
 \be \label{eq:redHerring} 
\VS \mathrm{\, is\,max} \quad \Longleftrightarrow \quad \QWPot(z) \propto \sin(2k_0z).
\ee
Putting back $|\PsiEnv (z)|^2$, which means taking into account the finite width of the quantum well, will induce slight distortion\footnote{An exact optimization reflecting such distortions has been done in Ref.~\cite{thayil_optimization_2025}.}, but one expects that the resulting optimum is close to the simple sine, ending the optimization analysis.

The foremost problem with this optimal solution is that it, so far, could not be realized with current epitaxy techniques. The required oscillation wavelength, $2\pi/2k_0=0.32\,\mathrm{nm}\approx0.59\,a_\mathrm{Si}$, is too short.
It basically requires control over the Ge concentration down to monolayers \cite{losert_practical_2023}, being far below the demonstrated resolution limit of roughly 1 nm \cite{Wuetz2022}.
Harmonic oscillations with longer wavelengths, combined with either the Umklapp process \cite{Feng2022} or strain \cite{Woods_strain_and_osc_2024, adelsberger_valley-free_2024}, couple the valleys, but such higher-order couplings are typically weak, and the additional components, such as engineered strain, increase fabrication complexity even if successfully realized.

The second issue with Eq.~\eqref{eq:redHerring} is subtler. On the one hand, Eq.~\eqref{eq:VSisFT} has been derived within the effective mass theory, and thus the functions in it, including $\QWPot$, are continuous. On the other hand, $\QWPot$ is used to describe effects that have atomic origins, such as sharp quantum well boundaries, or oscillations over distances of the order of the lattice constant. Fortunately, this apparent inconsistency is softened by the finding that the proportionality expressed in Eq.~\eqref{eq:VSisFT} holds beyond the effective mass framework.\footnote{Ref.~\cite{losert_practical_2023}, referring to an expression equivalent to our Eq.~\eqref{eq:VSisFT}: ``This is a powerful statement that transcends effective-mass theory. ...such a universal description of valley splitting is quantitatively accurate for all quantum well geometries studied here, including interface steps, broadened interfaces, alloy disorder, Wiggle Wells, and other phenomena.''} We build on this finding. 

\subsection{Reinterpreting the Fourier transform picture }

We adopt Eq.~\eqref{eq:VSisFT} and accept that, concerning the valley splitting, the single-atom effects dominate. Thus, the potential $\QWPot(z)$ in Eq.~\eqref{eq:VSisFT} is solely due to atoms, specifically, due to replacing some of the Si atoms by Ge. Accordingly, similar to \cite{Lima2023,lima_valley_2024}, we consider\footnote{What exactly is the parameter $\VGe$ is a subtle question, which we discuss in App.~\ref{app:E0}.} 
\be
\label{eq:discretizedQWPot}
\QWPot (z) = \sum_{m \in \mathrm{Ge}} \VGe \delta(z-z_m).
\ee
Here, $m$ are the labels of Ge atoms and $z_m$ are their positions. Particularly, there is neither an ``external electric field''  nor any ``heterostructure interface(s)'' in $\QWPot(z)$. The former, namely a Hamiltonian term such as $e E z$, does not generate appreciable valley splitting directly \cite{saraiva_intervalley_2011} (even though it enters indirectly, by setting the wave function profile and the weights $w_n$ defined below), while the latter is only an emerging effective description relevant for the effective-mass quantity $\PsiEnv(z)$; in Eq.~\eqref{eq:VSisFT} this effect is present implicitly through the 
$z$ dependence of the Ge density
defining $\QWPot(z)$.%

Inserting Eq.~\eqref{eq:discretizedQWPot} into Eq.~\eqref{eq:VSisFT}, we get 
\be
\label{eq:VSisSUM}
\VS = E_0 \left| \sum_{m\in \mathrm{Ge}} v_0 |\PsiEnv(z_m)|^2 e^{-2ik_0 z_m} \right|,
\ee
with $v_0$ the volume corresponding to one atom. We have also introduced an overall scale $E_0$, which incorporates the single-atom potential strength $\VGe$ and the prefactor omitted in Eq.~\eqref{eq:VSisFT}. We will treat $E_0$ as a fitting parameter. The valley splitting is now given as a sum of complex numbers of unit modulus, $e^{-2ik_0 z_m}$, that is, phases, weighted by the envelope function $|\PsiEnv(z_m)|^2$. 

Let us assume that one can place a fixed number $M$ of Ge atoms within the (so far, one-dimensional) quantum well, defined now as the region where $|\PsiEnv(z)|^2$ is appreciable. The configuration that gives the maximal sum in Eq.~\eqref{eq:VSisSUM} is clearly one in which all the terms have the same phase. We call such a configuration \textit{synchronized} or \textit{resonant} phases (of individual Ge atoms). If the phases are synchronized, the valley splitting is simply proportional to $M$
\begin{equation} \label{eq:VSisM0}
\VS \approx E_0 M/N.
\end{equation}
The minor effects of the shape of $|\PsiEnv(z)|^2$ can be either neglected here entirely (in the previous equation, we have replaced it by $1/N$ with $N$ the number of monolayers in the quantum well), or taken into account by including the weights $w_n \propto |\PsiEnv(z_n)|^2$, normalized such that $\sum_{n\in \mathrm{all}} w_n=1$, where $n$ runs through all atoms, both Ge and Si. For later convenience, we introduce $\rho_n$ as a notation to distinguish a ($n$-th) site occupied by Ge using $\rho_n=1$ and Si with $\rho_n=0$. With this, a more precise definition of the fraction of $M/N$ from Eq.~\eqref{eq:VSisM0} is
\begin{align} \label{eq:VSisM}
\VS = E_0 \underbrace{\left| \sum_{n} w_n \rho_n e^{-2ik_0 z_n} \right|}_{\MEFFsymbol}.
\end{align}

The weighted sum $\MEFFsymbol$ will be called the {\MEFFname}.\footnote{In analogy to basic solid-state theory (see, for example, Eq.~(3.16) in Ref.~\cite{ashcroft_solid_1976}) in which one assigns `geometrical {\MEFFname}s' to unit cells, molecules, and other atomic arrangements.}
It is a positive dimensionless quantity, bounded from above by 1. This value would be achieved in the hypothetical case of all sites $n$ being scattering sites, $\rho_n=1$, and all phases $e^{-2ik_0 z_n}$ being the same. These two conditions, however, are contradictory, and the achievable values for $\MEFFsymbol$ are smaller, as we will see below. 

In any case, beyond defining $\MEFFsymbol$, all other properties of $\QWPot(z)$ are irrelevant, \textit{including its periodicity}.
This observation concludes our theory of valley splitting and is central to all results that follow. 

The next essential fact is that the possible locations of the Ge atoms are restricted to a discrete grid, the crystal sites. In our one-dimensional model encompassed by Eq.~\eqref{eq:VSisSUM}, a single atom represents a two-dimensional monolayer of a three-dimensional structure. We keep this simplification for now and will remove it later. The available slots to place the Ge atoms then lie on a discrete one-dimensional grid, with the monolayer width being the grid step,
\be
\label{eq:discreteGrid}
z_m \in \left\{ \frac{a_\mathrm{Si}}{4} n \right\}_{n \in \mathrm{Integers}}. 
\ee
Only those sites of this grid that are occupied by a Ge atom (and not a Si atom) are in the set $m \in \mathrm{Ge}$. 

We now give two alternative interpretations of the result in Eq.~\eqref{eq:VSisM} based on simple physical considerations.  

\subsection{A resonant-scattering picture}

We first consider the valley splitting as a scattering problem. We assume that an electron starts in one of the valley states corresponding to the silicon band structure, say $\PsiFull_+$. We ignore the envelope function $\PsiEnv$ and take $\PsiFull_+$ to be a plane wave, modulated on short scales by the periodic function $u_+$. We consider how this state is backscattered\footnote{The name ``backscattering'' is slightly misleading, since here, in contrast to a typical scattering problem, the states $\PsiFull_\pm$ are not propagating. However, this is irrelevant for the discussion and Eq.~\eqref{eq:scatteringState}.} by ``scatterers'' that are created by replacing some of the Si atoms with Ge. Assuming that the scattering is weak, the exact scattering state is
\be
\label{eq:scatteringState}
|\PsiScattering_+ \rangle = | \PsiFull_+ \rangle  + \sum_{i \in \mathrm{scatterers}} \PsiFull_+(z_i) s_i e^{i \phi_i} |\PsiFull_-^{z_i}\rangle.
\ee
Here, $\PsiFull_+(z_i) \propto \exp(+ik_0z_i)$ is the complex amplitude of the incoming wave hitting the $i$-th scatterer, the real number $s_i$ is the scattering amplitude\footnote{Strictly speaking, $s_i$ should be infinitesimal, for the formula in Eq.~\eqref{eq:scatteringState} to be exact; but it is enough if it is small, $s_i M\ll1$.}, $\phi_i$ is the scattering phase, and the last term is the backscattered plane wave that starts off at position $z_i$, $\langle z|\PsiFull_-^{z_i}\rangle \propto \exp[-ik_0(z-z_i)]$. Since all scatterers are identical (Ge atoms), we have $s_i=s$ and $\phi_i=\phi$. These parameters depend on the microscopic form of the Ge atom potential $\VGe$. The scattering-matrix amplitude becomes
\be
S_{-+} = \langle \PsiFull_- | \PsiScattering_+\rangle = s e^{i\phi} \sum_{i}e^{-2ik_0 z_i}.
\ee
Assuming that, in this infinitesimal scattering scenario, the Hamiltonian and scattering off-diagonal matrix elements are proportional, we obtain Eq.~\eqref{eq:Delta} as $\Delta \propto S_{-+}$.
We get Eq.~\eqref{eq:VSisSUM} upon reinserting the neglected slow envelope modulation $\PsiEnv$. With the scattering picture, Eq.~\eqref{eq:VSisM} is very natural. Since the scatterers are identical, their interference being constructive or destructive is determined solely by their placement relative to the oscillating phase $\exp(-2ik_0z)$. If they are all placed such that the local phases $\exp(-2ik_0z_i)$ are identical, the backscattering is maximal (all scatterings in phase), and its amplitude is proportional to the number of scatterers. One would call this situation \textit{resonant scattering}.

\subsection{Variational energy-minimization picture}
\label{sec:variational}

Another approach is to formulate the valley splitting as a variational problem. Let us take a single electron at the conduction band edge of a silicon crystal. Replacing some of the Si atoms by Ge leads to a change in the Hamiltonian $\delta H $ and a new ground state. We search for it within the originally degenerate two-dimensional lowest-energy manifold, spanned by the basis states $\{ \PsiFull_+, \PsiFull_-\}$. Thus, the variational ansatz is
\be
\label{eq:ansatzVar}
| \PsiVariational \rangle = \alpha |\PsiFull_+ \rangle + \beta e^{-i\phi} |\PsiFull_- \rangle.
\ee
Here $\alpha$, $\beta$, and $\phi$ are three real numbers, the variational parameters. Their values are found by minimizing the variational energy 
\be
\delta E = \langle \PsiVariational | \delta H | \PsiVariational \rangle. 
\ee 
We approximate the $m$-th Ge atom placed at $z_m$ by a delta function potential, and set $\delta H = \QWPot$, with the latter as given in Eq.~\eqref{eq:discretizedQWPot}.
Anticipating the result, we choose $\alpha=\beta=1/\sqrt{2}$ and cast the variational energy as a function of the electron density $\rho(z)=|\PsiVariational(z)|^2$,
\be
\label{eq:variationalResult}
\delta E = \sum_{m \in \mathrm{Ge}} \VGe( \underbrace{1+\cos(2k_0 z_m + \phi)}_{\rho(z_m)} ).
\ee
Let us assume that the potential $\VGe$ is attractive. The largest possible energy drop $\delta E$ arises if the perturbing Ge atoms are placed such that, by tuning the phase $\phi$, the density can be adjusted so that every Ge atom falls on the local maximum of the density $\rho(z_m)=2$. The orthogonal state will correspond to density $1-\cos(2k_0 z + \phi)$, taking the value zero at the positions of Ge atoms. The variational energy of this state is zero, and the difference of the two energies, being $|\delta E|$, is the valley splitting. We again arrive at Eq.~\eqref{eq:VSisM}, which is now interpreted as a sum of $M$ individual contributions of variational energies $\VGe$. They are additive, resulting in maximal variational total energy gain, if and only if every Ge atom is at a local density maximum (or minimum if $\VGe$ is repulsive). Beyond this condition of \textit{local synchronization}, there is no requirement of periodicity in the placement of the Ge atoms.

\begin{figure}
	\centerline{\includegraphics[width=\linewidth]{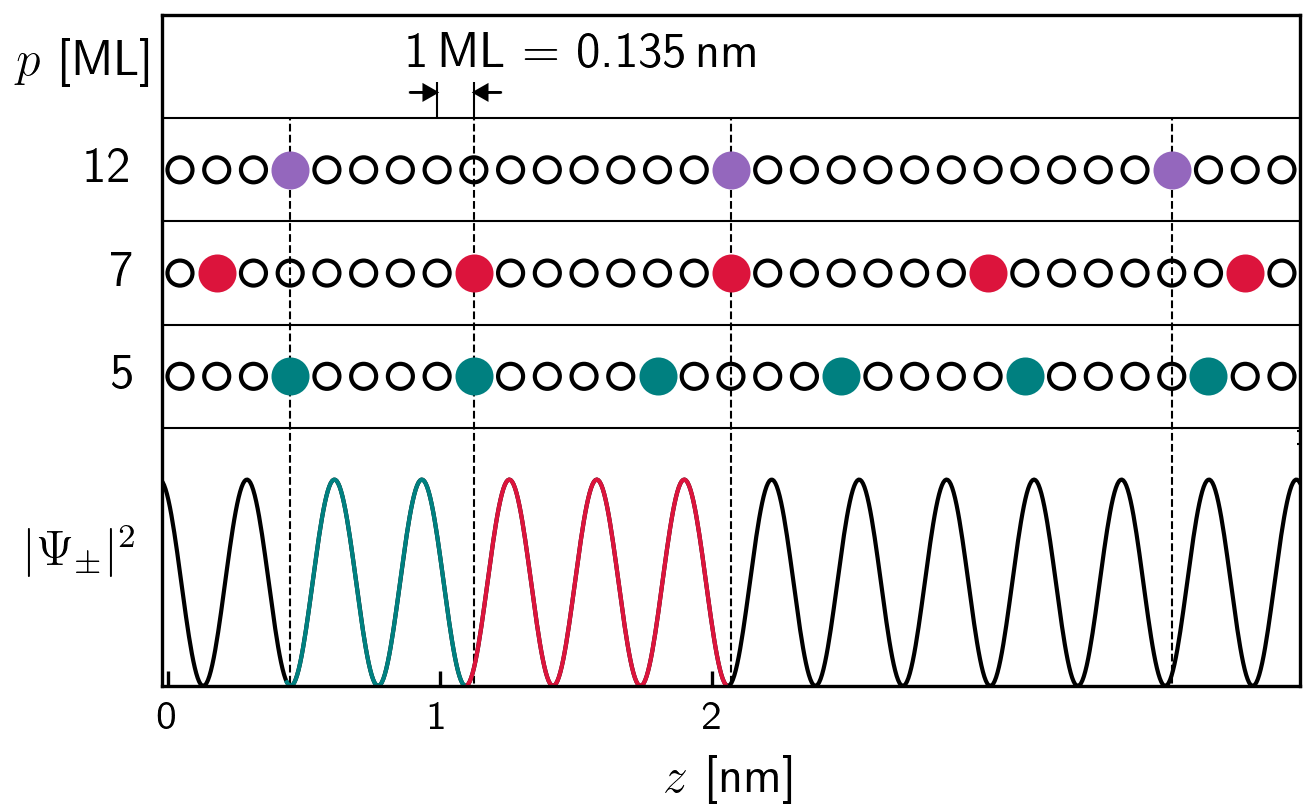}}
    \caption{Ge-doped monolayers (colored circles) with distances 12 (violet), 7(red), and 5(green) are approximately commensurate with the valley interference pattern $|\PsiFull_+(z) + \PsiFull_-(z)|^2 \propto  1 + \cos(2k_0z +\phi)$.}
    \label{fig:commensurability}
\end{figure}

\subsection{Incommensurability and the lucky coincidence}

As one would expect, in reality the Ge atoms can not be placed such that they are all perfectly \textit{resonant} or \textit{synchronized}. Apart from practical limitations on the resolution of heterostructure epitaxy, another reason is that the period of the oscillations in $\rho(z)$, being $2\pi/2k_0$, and the grid constant in Eq.~\eqref{eq:discreteGrid}, being $a_\mathrm{Si}/4$, are not commensurate. The silicon conduction band minimum is located inside the Brillouin zone, at a wave vector $k_0$ which has no special relation to the zone boundary. Nevertheless, plotting the density $\rho(z)$ together with the discrete grid in Fig.~\ref{fig:commensurability} we observe a lucky coincidence. The density oscillations almost perfectly match the grid positions 12 sites apart, say at $m=0$ and $m=12$, with only a small misalignment when inserting one additional Ge between them, either at $m=5$ or $m=7$. The grid spacing of 5, 7, and 12 monolayers corresponds to 0.675, 0.945, and 1.62 nm, respectively. With respect to practical fabrication, these numbers look much better than the $2\pi/2k_0\approx 0.32\,\mathrm{nm}$ required by Eq.~\eqref{eq:redHerring} and are feasible with state-of-the-art molecular beam epitaxy (MBE).

\subsection{Interpretation and conclusions on the valley-splitting theory}

We give two examples illustrating that the physical interpretations provided are useful. First, the variational picture gives an interpretation of the intervalley phase (which is a well-defined physical parameter with practical consequences \cite{tagliaferri_impact_2018}) as a simple variational parameter minimizing the ground state energy (see Ref.~\cite{Zwanenburg2013} page 12 for a related discussion). Second, the resonant scattering picture suggests that without any optimization of the Ge placement, the total accumulated complex backscattering amplitude will vary as $\sqrt{M}$. We show in App.~\ref{app:randomWalk} that the tight-binding data indeed follow this trend. As a trivial conclusion based on the properties of a simple random walk, this leads us to the quantitative prediction,
\be
\VS \approx E_0 \sqrt{0.41\frac{a_\mathrm{Si}^3}{16 \pi l_E l_x l_y} }\sqrt{\rho},
\ee
for the valley splitting due to a uniform Ge doping $\rho$ within a quantum well defined by a Si/SiGe interface and the electric field $E$, which relates to the length $l_E$ by $l_E^3 = \hbar^2/2m_l e E$. As is well known for a random walk, the statistical fluctuations are as large as the mean value itself, and thus a uniform doping can not deliver a valley splitting reliably offset from zero.

With these physical insights, we conclude the following. Accepting that the valley splitting, expressed in Eq.~\eqref{eq:VSisFT}, is due to atomic effects leads us to consider a model where $\QWPot$ is composed of discrete delta-like contributions. They represent selected sites on a discrete grid that are occupied by Ge, instead of Si atoms. On the one hand, this lets us relate the valley-splitting magnitude and phase to parameters representing a single-atom potential such as $\VGe$, through the physics of resonant scattering or variational minimization. More importantly, this structure of $\QWPot$ shows that Eq.~\eqref{eq:redHerring} is a red herring.\footnote{We revisit Eq.~\eqref{eq:redHerring} in App.~\ref{app:trouble}.} Such a harmonically oscillating profile is not within the definition domain of $\QWPot$. Figure \ref{fig:commensurability} shows that, instead, one should examine heterostructures with 5, 7, and 12 monolayer distances between Ge doping layers in various combinations.\footnote{We again point out Ref.~\cite{zhang_genetic_2013}, which also considered sharp doping with Ge layers, though not inside the silicon quantum well, but outside it, in the barriers. Interestingly, they found that the Bragg-mirror physics they aim at requires different magic numbers (all their best patterns start with a 4-monolayer-wide Ge segment, followed by alternating segments of Si or Ge with monolayer widths 2, 4, or 6).}

\subsection{Aspects of the real problem that were ignored so far}
\label{sec:real aspects}

We now consider aspects the simple model above ignored. First, the real problem is three-dimensional. We reflect it by promoting a single Ge atom in the one-dimensional model to represent a crystal atomic plane, a monolayer. With this change, the parameter $\rho_n$ denotes the Ge concentration in monolayer $n$. The scattering picture above implies that the atomic disorder within a single monolayer\footnote{By which we mean the fact that only a fraction of Si atoms within a given monolayer will be replaced by Ge. Monolayers with nominally identical Ge fractions will have different configurations of Ge atoms.} is irrelevant; all Ge atoms within a single monolayer are automatically resonant. Second, the Ge doping can not be controlled with monolayer selectivity. Aside from possible issues with the timing of switching the Ge source on and the homogeneity of the epitaxial growth, atomic diffusion will spread Ge atoms across neighboring monolayers. Third, in the above models, we have treated the valley splitting in the first-order perturbation theory, assuming that the atomic potential $\VGe$ is weak. This is not necessarily so. For example, in the resonant scattering picture, it is not hard to imagine multiple scattering effects to invalidate Eq.~\eqref{eq:scatteringState} in scenarios reminiscent of resonant transmission through a multi-barrier structure. 
To judge the fate of our central finding, Eq.~\eqref{eq:VSisM}, against all these departures from the toy model simplifications, we turn to numerics. 
 
 \subsection{Numerics models and methods}
 \label{sec:QW nomenclature}
 
We mostly use tight-binding methods, supplemented with some DFT calculations.  We model a SiGe/Si/SiGe heterostructure grown along the [001] direction. There are $N_b$ monolayers in the first SiGe buffer layer, $N$ in the Si quantum well, and $N_b$ again in the second buffer (a monolayer is perpendicular to the growth direction). The Si quantum well is laterally strained by 1.15\%. An electric field $E$ is applied across the quantum well. 

We investigate the valley splitting arising from additional Ge doping in the quantum well. We consider that this doping is defined by the Ge concentration $\rho_n\in[0,1]$ fixed for each monolayer $n$. Together, such concentrations define some (heterostructure) ``profile'', being the set $\{ \rho_n\}_{n=1}^{N_b+N+N_b}$.  In numerics, such a profile is realized by replacing $N_n^\mathrm{Ge}$ of the $N_n=N_n^\mathrm{Ge}+N_n^\mathrm{Si}$ atoms with Ge. The replaced atoms are selected randomly. In most of the figures, the integer $N_n^\mathrm{Ge}$ is fixed so that $N_n^\mathrm{Ge}/N_n \approx \rho_n$.\footnote{Thus, except for Fig.~\ref{fig:real}, our numerics do not include the random, binomial fluctuations of the Ge atom count due to the finite total number of atoms in a monolayer. In Fig.~\ref{fig:real}, where we include it, we find that it dominates the statistical spread of the valley splittings for a fixed doping profile, in line with previous findings \cite{Cvitkovich_VS_2}. In App.~\ref{app:atomCountFluctuations} we investigate the effect on the model-data correlation over a large dataset.} The pristine SiGe/Si/SiGe heterostructure without additional Ge doping is represented by taking $\rho_n=0$ in the central Si region and finite $\rho_n$, typically 30\%, in the SiGe buffers, with a transition region of a few monolayers where the density connects the two values smoothly.

For the tight binding, we use an $sp^3d^5s^*$ model~\cite{Niquet2009} in conjunction with the Keating valence force field~\cite{VFF} for structural relaxation. Spin-orbit coupling is included in the tight-binding Hamiltonian.\footnote{Interestingly, sharp atomic-scale Ge concentration profiles have recently been shown to also enhance the intrinsic (Dresselhaus) spin-orbit coupling, through a Fourier component at a different reciprocal wavevector than the one responsible for valley splitting~\cite{soomro_high-fidelity_2026}.} We consider heterostructures with no planar confinement, with a supercell size of $16\times16$ unit cells in the plane and $200$ monolayers in the growth direction, including two SiGe buffers with $N_b=64$ and $N=72$ monolayers for the middle Si region. The total system consists of about 10$^5$ atoms. We have verified earlier that this simulation cell size is well converged with respect to the median values of the valley-splitting~\cite{Gradwohl2024}. Random sampling of Ge configurations indeed results in valley splittings that have statistical fluctuations. While our cell size is small relative to a typical quantum dot -- where the effects of such statistical fluctuations would be smoothed out by spatial averaging\footnote{This effect can be seen, for example, in Fig.~S4 in the supplementary material of Ref.~\cite{Gradwohl2024}.} -- the fluctuations in our numerical results are artificially constrained. This is because (except for Fig.~\ref{fig:real}) we maintain a constant number of Ge atoms per monolayer, thereby neglecting the stochastic variance in dopant count that would occur in a physical realization (see App.~\ref{app:atomCountFluctuations}). In a realistic device, these two effects -- spatial averaging and stochastic dopant distribution -- will compete.

For DFT calculations, we employ the machinery of Ref.~\cite{Cvitkovich_VS_2}, specifically the \texttt{CP2K} code~\cite{CP2K}, which relies on atom-centered basis sets, allowing for efficient simulation of large systems. (Unlike tight-binding, our DFT code does not include spin-orbit interactions). All calculations are performed with the hybrid functional PBE0~\cite{PBE0_TC_LRC}, which is known to give accurate band gaps in semiconductor materials~\cite{Functionals2011}. The orbitals are expanded in double-$\zeta$ basis sets, and the core electrons are approximated by Goedecker-Teter-Hutter (GTH) pseudopotentials~\cite{GTH}. The DFT supercell contains 136 layers in the growth direction and 18 atoms per layer (2448 atoms in total). For further details, please see Ref.~\cite{Cvitkovich_VS_2}.

\section{Valley splitting from numerics for monolayer-thick spikes}

To examine the trends that different doping structures imply for the valley splitting, we first consider unrealistic profiles where $\rho_n=0$ for most Si-region monolayers, except for a small fraction of selected ones where $\rho_m= \mathrm{const}$\footnote{As typical values, we use $\rho_m=10\%$ in Fig.~\ref{fig:ideal} and \ref{fig:survey}, 5.5\% in Fig.~\ref{fig:DFT}, and up to 7.5\% in Fig.~\ref{fig:real}.} That is, we first consider (delta) doping in one-monolayer thin spikes. Later, we consider realistic profiles, where such delta-like spikes are broadened according to what is observed in STM structural characterizations of experimental samples. 

\subsection{The magical integers 5 and 7}

\begin{figure}
	\centerline{\includegraphics[width=\linewidth]{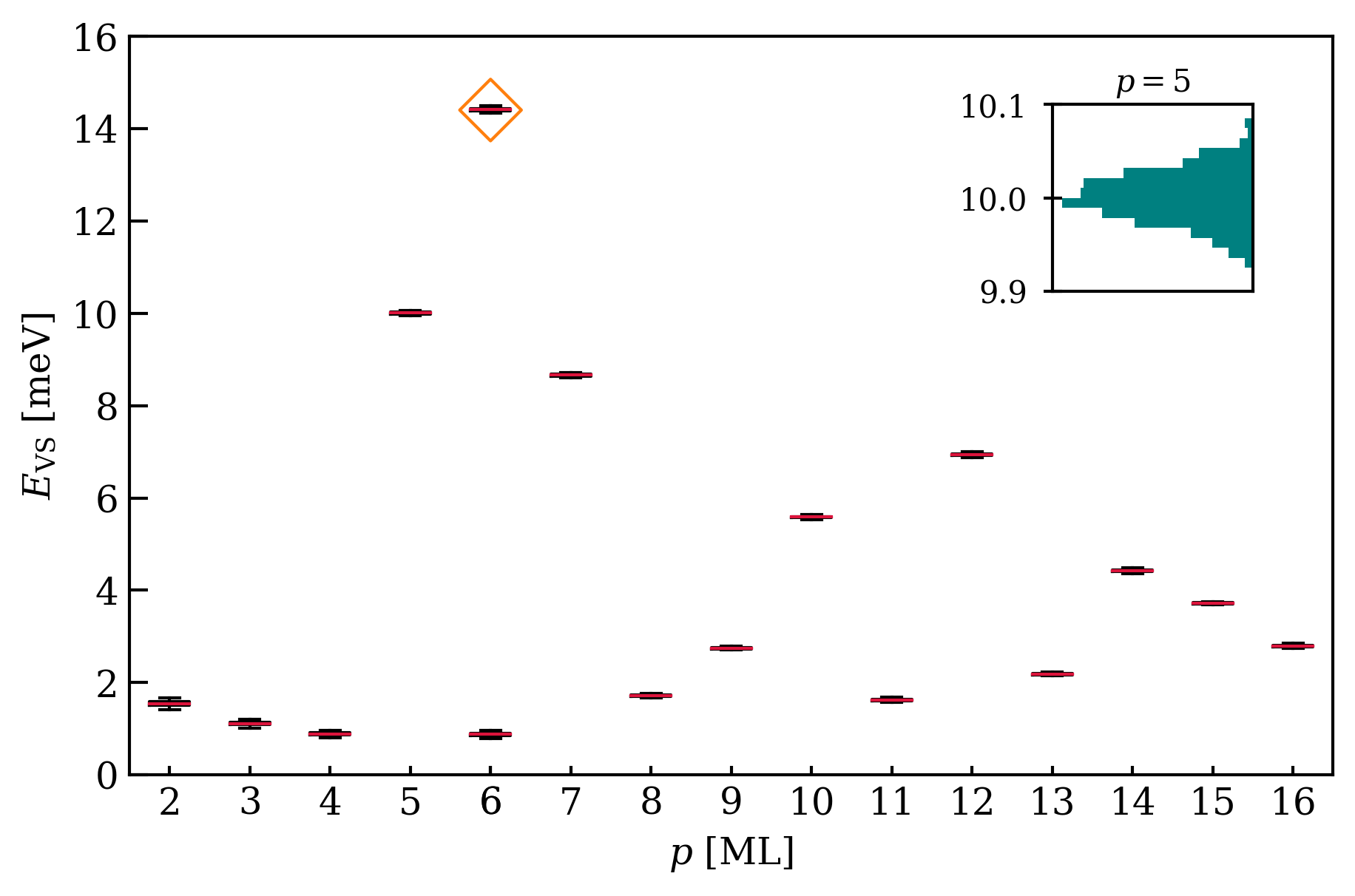}}
    \caption{
    Valley splitting as obtained from tight binding for delta-like Ge spikes (single Ge-doped monolayers) with a Ge concentration of $10\,\%$ separated by $p-1$ pure Si layers (except for the value in the orange diamond, which alternates 5 and 7 monolayer separations). For each $p$, we evaluate 100 realizations of the heterostructure with nominally identical doping profiles and randomly distributed Ge atoms. The statistical ensemble of the resulting valley splitting is depicted by its median value (red lines; the median is more robust to outliers than the mean, though for these narrow distributions the difference is immaterial) and its maximum and minimum (error bars; mostly too small to be clearly visible at the figure resolution).
    The electric field is $E=5\,\mathrm{mV/nm}$. 
    The inset shows the histogram of values for $p=5$.
    }
    \label{fig:ideal}
\end{figure}

We first demonstrate the importance of the coincidental resonances illustrated in Fig.~\ref{fig:commensurability}. To this end, we consider periodic profiles, in which $\rho_m$ is nonzero for $m$ being a multiple of an integer $p$. The resulting valley splittings are plotted in Fig.~\ref{fig:ideal}. The values for periods 5, 7, and 12 stand out as exceptionally large, just as suggested by Fig.~\ref{fig:commensurability}. Actually, the implications go further. Ordering the periods according to the resulting valley splitting, one gets the progression 5, 7, $12=5+7$, $10=5+5$, $14=7+7$, and $15=5+5+5$ as the first six entries, before encountering the first case (16) that has no simple relation to the two ``magical integers'' 5 and 7. 

Seeing the power of 5, 7, and their sum 12, it is hard to resist trying to combine them in synergy. We examine a structure\footnote{Its variants presented in App.~\ref{app:finalSuggestion} are our final suggestions for simple structures aimed at maximal valley splitting.} that does not have a simple periodicity, but alternates doping in layers displaced by units of 5-7-5-7-5-7$\cdots$. We find the valley splitting of 14.4 meV (the diamond point in Fig.~~\ref{fig:ideal}). This value, corresponding to a very simple doping profile, boosts the splitting compared to 6-6-6-6-6-6$\cdots$ that has the same Ge content, by a factor of 20. It beats all simple periods in Fig.~\ref{fig:ideal} and is very close to the highest values that we could reach by a non-trivial optimization (see below; especially Fig.~\ref{fig:survey} black points and App.~\ref{app:optimization}). It is also comparable to valley splitting observed in numerics in Ref.~\cite{losert_practical_2023} for complex profiles found using a different sophisticated optimization.
This is a striking demonstration of why it is useful to start with Eq.~\eqref{eq:VSisSUM} instead of Eq.~\eqref{eq:redHerring}.

\begin{figure}[h]
	\centerline{\includegraphics[width=\linewidth]{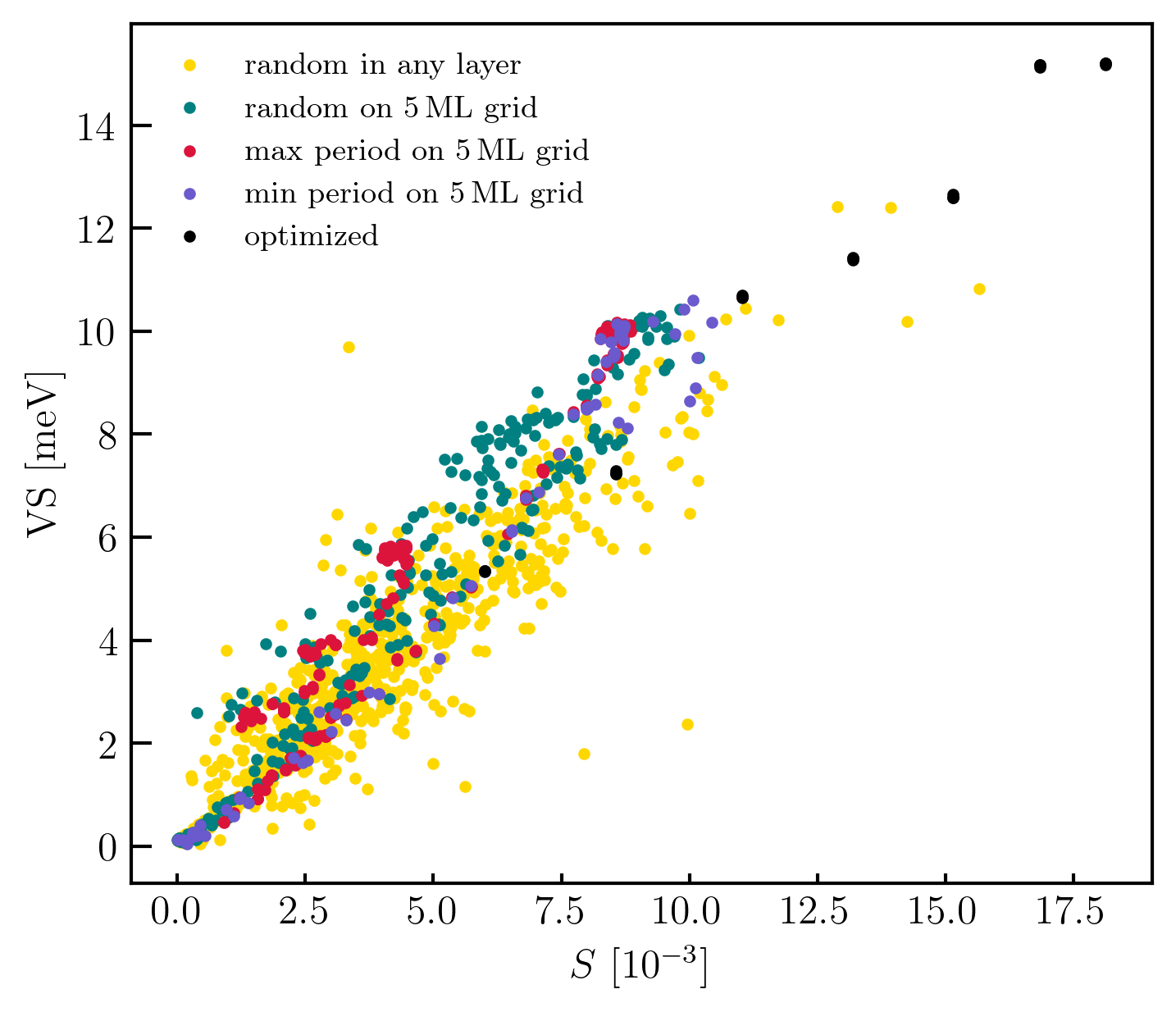}}
    \caption{
           \label{fig:survey}
        Valley splitting as a function of $\MEFFsymbol$ for various profile types. In all cases, the profile contains $M$ spikes, where $M$ ranges from 1 to 14, and we sample 20 structures (except for for yellow with 49 structures; and for black with 1 structure). In the selected monolayers, the Ge doping is $\rho_m=10\%$. %
      The colors mean the following.
     \textit{Yellow:} the spikes are distributed completely randomly (but not on top of each other).
    \textit{Black:} after fixing $M$, we obtain the profile by maximizing $\MEFFsymbol$ numerically (see App.~\ref{app:optimization} for the optimization algorithm).
    The remaining structures are constrained such that Ge spikes can only fall onto every 5-th Si monolayer ($p=5$ in Fig.~\ref{fig:commensurability}).
    \textit{Green:} random on the $p=5$ lattice.
    \textit{Red:} periodic (with period $q$) on the $p=5$ lattice. The period $q$ is taken as the largest integer multiple of $p$ that fits $M$ spikes into the quantum well. There remains one degree of freedom, a shift of the whole arrangement, which is chosen at random.
    \textit{Purple:} periodic (with the period $q=p$) on the $p=5$ lattice. Again, the arrangement is randomly shifted around in the well.
    }
\end{figure}

\subsection{The profile periodicity is irrelevant}

The enhancements observed in Fig.~\ref{fig:ideal} thus confirm the simple observations made in Fig.~\ref{fig:commensurability} in an impressive way.
Nevertheless, Fig.~\ref{fig:ideal} relies on periodic profiles.
We next inspect our central claim, Eq.~\eqref{eq:VSisM},\footnote{In line with adopting the lowest-order perturbation theory by Eq.~\eqref{eq:Delta}, we use $\Psi_\mathrm{env}(z)$ derived for a clean, undoped, quantum well here and throughout the article whenever we evaluate the {\MEFFname} using Eq.~\eqref{eq:VSisM}.} which states that the doping periodicity is immaterial. To this end, we examined a large number of profile configurations, including both periodic and non-periodic ones. The resulting valley splittings are gathered in Fig.~\ref{fig:survey}. 

First of all, the figure shows a clear correlation between the valley splitting $\VS$ and the {\MEFFname} $\MEFFsymbol$. However, more striking is that beyond that correlation, the periodicity plays no role. Once $\MEFFsymbol$ is fixed, there is essentially no difference between a profile with Ge-doping monolayers chosen completely randomly (yellow), with a partial periodic structure (green), with full periodicity (red and violet), or with a non-trivial non-periodic structure obtained by optimization (black). All these variants in the choice of how to build a doping profile have similar regression slopes (see App.~\ref{app:fitting} for a fit) and variations on top of them. We observe that different variants do impose further structure to the plot. For example, the yellow points have a larger variance than points of other colors, the violet points lie on an arch, the green ones start on a line with a small slope, and so on. We deem these effects of minor importance and do not pursue them further. By far the most important message of the figure is that the profile periodicity of any kind, including the $2\pi/2k_0$ one, is by itself irrelevant for the valley splitting. Instead, what matters is the total amplitude of the scattering on Ge impurities, expressed by the quantum well {\MEFFname} $\MEFFsymbol$. 

\subsection{The effects beyond Eq.~\eqref{eq:VSisM}}

Though the correlation between $\VS$ and $\MEFFsymbol$ is clear, so is the fact that the data in Fig.~\ref{fig:survey} are scattered and do not fall on a straight line, even approximately. We have anticipated such a possibility by pointing out that the ``linear response regime'' expressed by Eq.~\eqref{eq:VSisM} is expected only if the scattering on a Ge atom is weak. The substantial spread observed in the data suggests that this might not be the case. 

We first comment on how the alignment between $\VS$ and its prediction through $\MEFFsymbol$ can be improved within our simple approach. For the data plotted in Fig.~\ref{fig:survey}, the correlation coefficient between the abscissa and ordinate variables is 6.8\% smaller than 100\%. This discrepancy can be decreased substantially, to 1.4\%,\footnote{Including all the data in Fig.~\ref{fig:survey} irrespective of the color gives the correlation coefficient of 98.6\%.} 
by using a smaller $k_0 \approx 0.83$, the position of the conduction band minimum in the tight-binding band structure of bulk silicon biaxially strained onto Si$_{0.7}$Ge$_{0.3}$ (versus $k_0 = 0.85$ in unstrained silicon). This reveals that the shift of the conduction band minimum with strains (further discussed in Appendix \ref{app:strain}) leaves quantifiable fingerprints on the valley splittings.
The method to fit $k_0$ is described in App.~\ref{app:fitting} and related checks of the validity of Eq.~\eqref{eq:VSisM} are in Apps.~\ref{app:fittedParameters} and \ref{app:refinement}.

We suspect that the 98--99\% correlation between the exact numerics data and simple models such as Eq.~\eqref{eq:VSisMrefined} cannot be improved much further within the perturbative approach~\cite{Cvitkovich_VS_2}.\footnote{In Fig.~1(f) of Ref.~\cite{losert_practical_2023}, one can see data analogous to our Fig.~\ref{fig:survey} with a much higher correlation, very close to 100\%. We believe that the difference is due to a much larger computational cell adopted in Ref.~\cite{losert_practical_2023}, $100 \times 100$ nm$^2$ compared to our 8.7$\times$8.7 nm$^2$ cell. In a larger cell, the tight-binding output is effectively statistically averaged, suppressing data variance. In addition, Ref.~\cite{losert_practical_2023} presents data for lower Ge doping concentrations, which probably diminishes the statistical fluctuations further \cite{Cvitkovich_VS_2}. Finally, we mention that fluctuations in Ge atom count, an effect not included in our Fig.~\ref{fig:survey}, but present in real samples, will also increase statistical fluctuations and thus decrease the correlation coefficient; see App.~\ref{app:atomCountFluctuations}.}
The direct numerical support for this expectation is given in App.~\ref{sec:nonlinearity}.
To shed light on the effects that influence valley splitting in nanostructures, where the lattice and crystal potential are both strongly affected by Ge doping, we resort to DFT.

\begin{figure}[!hbtp]
        \centerline{\includegraphics[width=\linewidth]{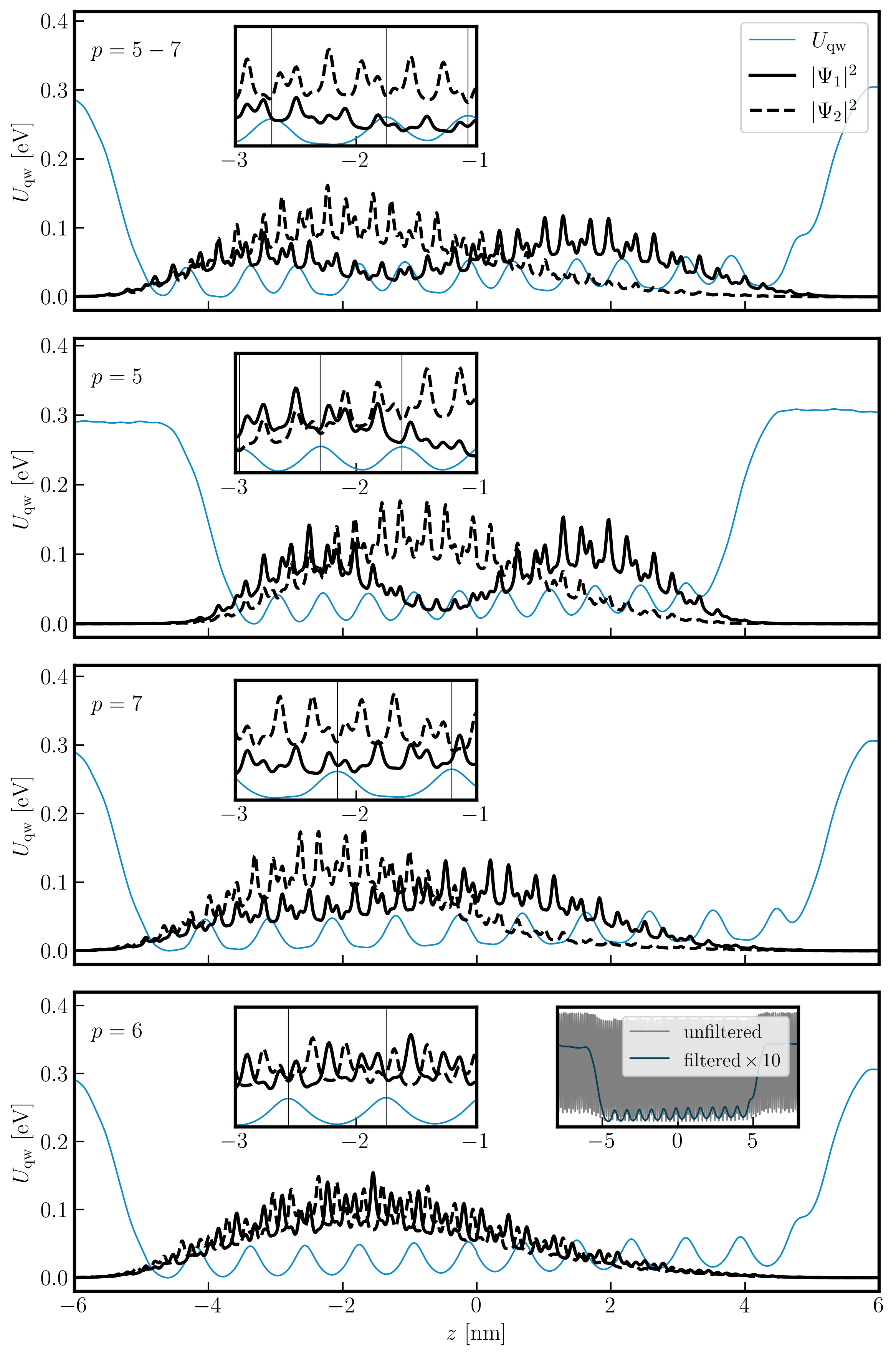}}
    \caption{
    \textbf{DFT results for a profile with Ge-doping spikes placed periodically at every $p$-th monolayer.} The blue curve shows the quantum well potential $U_\mathrm{qw}$ obtained by convoluting the Hartree potential with a Gaussian filter with a $0.25$ nm width. In the insets, the positions of three Ge monolayers are denoted by vertical lines. The unfiltered potential is shown in gray in the upper-right inset of the $p=6$ panel. The black solid (dashed) curve shows the electron density of the lowest (second-lowest) valley state, $|\Psi_{1,2}|^2$. Because of the limited supercell size (18 atoms per plane), the smallest nonzero Ge content for a monolayer is 5.5\%, which we used for the doped monolayers inside the quantum well. The barrier monolayers contain 5 Ge atoms (27.8\%).
 The DFT valley splittings are 8.38, 18.97, 2.10, and 0.68 meV for the $5$--$7$, $5$, $7$, and $6$ patterns, respectively. 
     \label{fig:DFT}
   }
\end{figure}

As an arbitrary example, we consider periodic doping patterns with periods $p=5,6$, and $7$ monolayers, as well as the alternating $5$--$7$ pattern. Figure~\ref{fig:DFT} shows some of the DFT output. The blue curves depict the Hartree potential. Moving from left to right, one can see the transition from the barrier region\footnote{The potential drops by about 300\,meV. This drop is not the conduction-band offset, since the Hartree potential does not include the non-local exact-exchange contribution of the hybrid functional. A better estimate can be drawn from the energy difference between the lowest bound state in the quantum well and the lowest unbound state in the barrier~\cite{Cvitkovich_VS_2}, giving 225\,meV, closer to the measured range 160--190\,meV~\cite{maiti_strained-si_1998}. 
}
to the quantum well region, where the potential drops to approximately zero meV, with further growth due to the electric field (causing the linear increase within the well).
On top of this smooth background set by the barrier, quantum well, and the electric field, we observe peaks at the positions of the Ge layers (denoted by vertical lines in the insets), confirming that $\VGe$ is repulsive. However, we note that the width of an individual peak is dominated by the smoothing filter used to reveal trends present on larger length scales. The unfiltered atomic potential (shown in the upper-right inset of the $p=6$ panel) displays strong oscillations on atomic length scales.

With this setting, we now look at the wave functions. The black solid (dashed) curve shows the probability density of the ground state (excited state). We observe that the densities do not oscillate at a single wave vector, and that they both develop local minima at the repulsive Ge monolayer planes. 
Both features lie beyond the variational ansatz of Eq.~\eqref{eq:ansatzVar}, which confines the wave functions to a two-dimensional manifold so that only one of the two states can be depleted at the Ge planes. The local depletion of both states reflects the deformation of the wave functions by a strong, repulsive potential, an effect beyond the first-order perturbation theory. Despite the difference, we find that Eq.~\eqref{eq:ansatzVar} is a good basis for the valley splitting model embodied in Eq.~\eqref{eq:VSisM}.

The alignment of the wave function minima with Ge monolayers 
inspires us to consider a simple extension of Eq.~\eqref{eq:VSisM} into a variational ansatz that would allow for such an alignment. The idea is to allow a difference between the crystal momentum $k$ of the wave functions in our variational ansatz and the conduction band minimum $k_0$, promoting $k$ to a variational parameter. Moving $k$ away from $k_0$ incurs a kinetic-energy cost but allows variational energy gain from better alignment of the wave functions' extrema with the Ge scatterers. We analyze this idea in App.~\ref{app:refinement}. Even though the increase in correlation is marginal, the same approach, meaning a simple extension of the variational ansatz, allows us to examine additional important aspects. In App.~\ref{app:strain}, we estimate the expected increase in valley splitting due to engineered strain, through tuning the value of $k_0$. We find that strain is not expected to provide drastic improvements. In App.~\ref {app:step}, we analyze how monoatomic steps in the quantum well affect valley splitting, an issue that has been extensively studied.  We conclude that monoatomic steps do not threaten valley splittings whose bare value (that is, the value without any step) is within the meV scale.

Overall, we interpret the DFT results as confirming that scattering effects are strong (not perturbative) in nanostructures amenable to DFT. 
The DFT valley splittings are much larger for periods $p=5$, 7 (and for the 5-7 pattern) than for $p=6$ monolayers, as in tight binding. Yet the ranking of best structures is different. The value of $k_0$ is not, however, exactly the same in DFT and tight binding (among other differences) \cite{Cvitkovich_VS_2}. Also, the wave functions of Fig.~\ref{fig:DFT} (except the $p=6$ panel) display significant valley-orbit mixing (different envelopes for the two valleys) due to avoided crossings with the valley states of higher orbitals at large valley splittings. Valley-orbit mixing is also prevalent, to some extent, in tight-binding calculations. We will not pursue the comparison between DFT and tight binding further, as it would require aligning various quantities ($k_0$, conduction band offset, electric field and relaxed geometries) as has been investigated at length in Ref.~\cite{Cvitkovich_VS_2}.
We conclude that there is little prospect of continuing to develop the theory along the lines that lead us to Eq.~\eqref{eq:VSisMrefined}; we conjecture that the 2\% discrepancy that we observe between its predictions and the exact numerics is dominated by non-perturbative effects, supported by findings presented in App.~\ref{sec:nonlinearity}. The exact numerics is thus necessary for a quantitatively reliable prediction of the valley splitting for a given configuration of Ge scatterers. 
On the other hand, the details uncovered by such exact numerics preclude simple insights into trends: it would be difficult to arrive at formulas such as Eq.~\eqref{eq:VSisM} starting from data such as shown in Fig.~\ref{fig:DFT}. The combination of exact numerics embodied by DFT and tight binding with simple analytics in the spirit of Eq.~\eqref{eq:VSisM} seems to be the right blend of quantitative and qualitative theories of the valley splitting.

\section{Valley splitting from numerics for realistic doping profiles}

So far, for simplicity in both the analysis and the numerical presentations, we have treated Ge doping as singular, one-monolayer-thick spikes. We now remove this limitation. We will test the configurations we identified as most promising, incorporating profiles we deem realistic to implement with current MBE technology (noting its recent progress in this direction \cite{rahlff_sharp_2026}). We set the latter according to the following considerations.

Even if the intended change in the doping concentration in a quantum well is discrete (discontinuous), it will not be implemented perfectly, and the resulting doping profile will be smeared. A common phenomenology is to fit the measured isotope concentration to a sigmoid function with parameter $\tau$ and assign ``the interface width'' as $4\tau$. Beyond the ambiguity in the interface-width definition, determining the width (or steepness or sharpness) of the transition region (or ``interface'') is challenging because of uncertainties in the actual concentration density. Pulsed laser atom probe tomography (APT)~\cite{Koelling_APT_2009, Dyck_APT_2017, Klos_2024, Wuetz2022} offers sub-nanometer resolution and 10 ppm sensitivity for detecting dopants and interface segregation in semiconductors. However, its results for interfaces might get distorted by reconstruction artifacts. There are correction algorithms for these, for example, the ``z-redistribution algorithm'', which aligns APT results with STEM measurements \cite{Dyck_APT_2017}, or the SIMS technique \cite{Koelling_APT_2009, Dyck_APT_2017, Klos_2024, Tröger_ToF_SIMS_2025}. The resulting profile estimation can be quite sensitive to the chosen method (see Fig.~2 in Ref.~\cite{Klos_2024} for an example). One also notes that the steepness of the transition is different for a decreasing and increasing Ge concentration,\footnote{Ref.~\cite{Klos_2024} comments this effect by: ``We attribute this interface broadening to the segregation of Ge atoms, expected for a SiGe growth front overgrowing Si (here $^{28}$Si) in ultra high vacuum epitaxy, termed as leading edge in the literature.~\cite{Kube_Ge_seg_2010, Godbey_Ge_1992, Fukatsu_Ge_1991}''.} it depends strongly on the growth temperature~\cite{Gradwohl2024}, and might be reduced by annealing~\cite{Klos_2024}.

After all these warnings, we now give some numbers that one can find in the literature. In Ref.~\cite{Dyck_APT_2017}, the widths of both interfaces of a SiGe/Si/SiGe quantum well were estimated between 1 and 2 nm (the spread is due to strong dependence on the evaporation direction). Ref.~\cite{Wuetz2022} gives 0.8 nm (six monolayers). Among the most recent works, Ref.~\cite{Gradwohl2024} claims that Ge gradients as big as 30\% per nm have been realized, and Ref.~\cite{Klos_2024} estimates that in their wafer the SiGe to Si transition happens over 2.4 monolayers. 

\begin{figure}
	\centerline{\includegraphics[width=\linewidth]{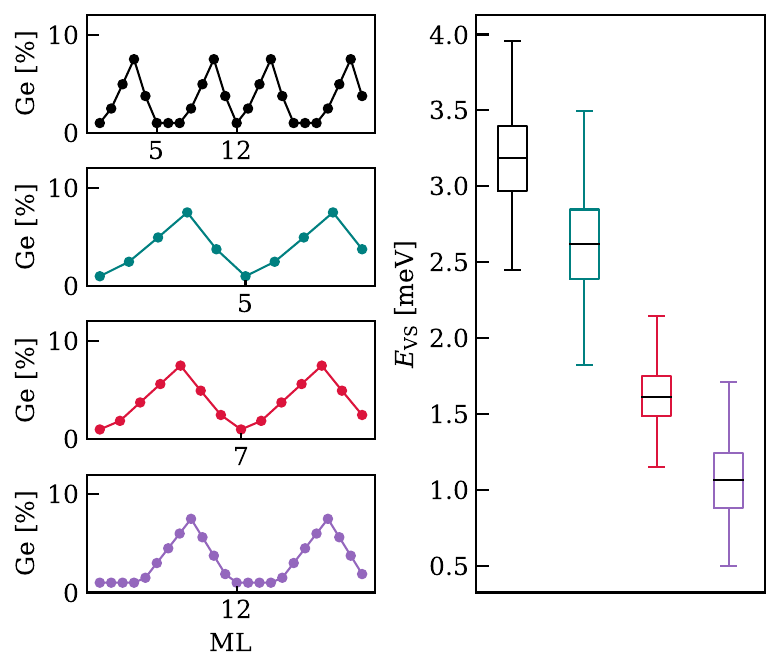}}
    \caption{
    \textbf{Tight-binding results for realistic heterostructures.}
    We analyze four instances of a quantum well built by repeating one of the blocks plotted on the left over the total width of 72 monolayers (about 10 nm) sandwiched by Si$_{0.7}$Ge$_{0.3}$ barriers. The four instances include three periodic spikes (5, 7, and 12 monolayer steps apart; the lower three panels on the left) and one more complex pattern, built from a 7-5-5-7 block (the top panel on the left). The electric field is $E=5\,\mathrm{mV/nm}$.
    The resulting valley splitting $E_\mathrm{VS}$ is shown in the right panel in corresponding colors. The IQR (interquartile range) is shown by a box, the extreme outliers by the bars, and the median is in black. 
    Compared to the values for ideal single-layer dopings, given in Fig.~\ref{fig:ideal}, the valley splittings are suppressed by factors of 0.22, 0.26, 0.19, and 0.15, respectively, for the four profile choices plotted.  These numbers can be explained by Eq.~\eqref{eq:suppression} which, using, $k_0 \approx 0.83$, gives $\alpha_{3,2}=0.26$,  $\alpha_{3,3}=0.17$,  $\alpha_{4,3}=0.24$, $\alpha_{4,4}=0.21$, and  $\alpha_{5,4}=0.14$, for a few profiles similar to the chosen ones.
    }
    \label{fig:real}
\end{figure}

Motivated by these recent advancements, we chose profiles in which a delta-like spike (a Ge-doped monolayer) is broadened into an approximately triangular shape with 3-5 monolayers on the side where the Ge concentration is increasing and somewhat less, 2-4, monolayers on the side where it is decreasing. In addition, we assume that entirely Ge-free monolayers are not feasible and set the residual Ge concentration to 1\,\%, which slightly distorts the triangular shape. %
The profiles are given in the left panel of Fig.~\ref{fig:real}. Unlike in the previous figures, the Ge atom count in each doped monolayer $N_n^\mathrm{Ge}$ is here sampled from a binomial distribution $B(N_n,\rho_n)$, so that the spreads in Fig.~\ref{fig:real} include the atom-count fluctuations. The most important observation is that the numerics predict valley splittings in the meV range upon the doping peak half-width broadened to up to 4 monolayers, well beyond the number 2.4 quoted in Ref.~\cite{Klos_2024}. 

We can also be more quantitative. If we assume that different peaks are approximately well resonant, the smearing due to a finite width is due to replacing a single complex number $e^{-2ik_0 z_n}$ with modulus one by a sum 
\be
\label{eq:suppression}
\alpha_{l,r} = \left| \sum_{n} \rho_{l,r}(n) e^{-2ik_0 z_n} \right|,
\ee
where the weights $\rho_{l,r}(n)$ are proportional to the Ge concentrations along the smeared peak, reaching $\rho(n_0)=1$ at the peak maximum. The notation reflects our choice of approximately triangular profiles, where these weights increase from 1\,\% to the target Ge concentration over $l$ monolayers and then drop back to 1\,\% over $r$ monolayers. With this definition, the suppression factor of an ideal spike is $\alpha_{1,1}=1$ (thus, no suppression), and for a broader peak, it is smaller. The four median values plotted in the right panel of Fig.~\ref{fig:real} are in acceptable correspondence with Eq.~\eqref{eq:suppression}~\footnote{The fluctuations, which in Fig.~\ref{fig:real} are dominated by the atom-count statistical fluctuations, will be smaller in a quantum dot of a realistic size due to averaging effects (the fluctuations decay as $1/R$ with $R$ the dot radius). The scaling with $R$ can also distinguish the stochastic and deterministic regime: In the former, the valley splitting median decays as $1/R$, and in the latter, it is independent of it. The independence of $R$ in the deterministic regime is a test alternative to observing the median being much larger than fluctuations. 
}.

Abandoning the monolayer-thin Ge scattering planes in favor of realistically smeared doping profiles decreases the predicted valley splitting by a factor of 4-6. 
Taking this factor into account for the values predicted in Fig.~\ref{fig:survey}, Fig.~\ref{fig:visualization}, and Tab.~\ref{tab:best}, we believe that Si/SiGe wafers with valley splittings in the meV range can be realized.

\begin{center}
\textbf{Conclusions}
\end{center}

\begin{table}
 \begin{center}
\begin{tabular}{c@{\quad}c@{\quad}c@{\quad}cc}
\toprule
$E$-field & QW & $k_0$ & \multirow{ 2}{*}{best doping pattern}\\
$[\mathrm{V}/\mu\mathrm{m}]$ & [nm] & [2$\pi/a_\mathrm{Si}$] & \\
\midrule
5 & 11.5 & 0.83 & x-5-5-7-5-5-7-5-7-5-5$\cdots$\\
1 & 11.5 & 0.83 & x-5-7-5-5-7-5-7-5-5-7-5-7-5$\cdots$\\
5 & 11.5 & 0.85 & x-5-4-5-5-7-5-4-5-5-5-5-5-5$\cdots$\\
1 & 11.5 & 0.85 & x-7-7-7-5-7-7-7-7-5-7-7-7$\cdots$\\
\bottomrule
\end{tabular}
\end{center}
\caption{\label{tab:best}
\textbf{Best patterns for selected sets of problem parameters.} The table collects the patterns given in Fig.~\ref{fig:visualization}. The pattern is given as the distance between consecutive Ge monolayers, starting at the SiGe/Si interface against which the electric field pushes the electron. The pattern specification is terminated at a point beyond which the influence is negligible, defined as contributions that change $\MEFFsymbol$ by less than $10^{-3}$.
The first entry, denoted by x, is not critical and can be set to any suitable small integer. Instead of a distance, it denotes the start of the doping pattern. In Fig.~\ref{fig:visualization}, these reference labels are 1, 5, 4, and 2 for the four given patterns, respectively, and refer to the monolayer indices of the barrier (the nominally undoped Si quantum well starts at monolayer index 6).}
\end{table}

We have theoretically revisited the problem of conduction-band valley splitting in Si/SiGe quantum wells, using tight-binding, DFT, and effective-mass models. We propose a novel view in which the Ge dopants are considered as scatterers on a discrete lattice. This view leads us to examine quantum well profiles that have not been suggested before, and do not directly relate to the oscillation period $2\pi/2k_0$. Instead, they are built by placing Ge-doping layers at integer monolayer distances (mostly 5 and 7, and their combinations). This view provides insight into the physics of valley splitting and enables simple optimization of quantum well profiles to enhance it. We examined various aspects of the problem, and particularly the sensitivity of the valley splitting to key parameters (electric field, quantum-well interface, strain, conduction-band minimum, electron mass). 
We have also investigated the role of monoatomic steps, and did not find them threatening to the proposal. Our best suggestions for optimal quantum-well profiles are given in Tab.~\ref{tab:best}, and we believe that based on them, one could push the valley splitting into the meV range with the present MBE technology. 

\begin{center}
\textbf{Acknowledgments}
\end{center}
This work was funded in part by the German Research Foundation (DFG) within the projects 289786932 (BO 3140/4-2), 443416183 (SPP 2244) and 314695032 (SFB1277) and by the EU 2DSPIN-TECH Program (Graphene Flagship). This work was also supported in part by NCCR SPIN, a National Center of Competence in Research funded by the Swiss National Science Foundation (grant number 225153). This publication is based on work supported by King Fahd University of Petroleum \& Minerals. The author at KFUPM acknowledges the Deanship of Research and the Center for Advanced Quantum Computing for the support received under Grant no. CUP25102 and no. INQC2500, respectively. YMN acknowledges support from the ``France 2030'' program (PEPR PRESQUILE-ANR-22-PETQ-0002), and from the Horizon Europe Framework Program (grant agreement 101174557 QLSI2). This project was funded within the QuantERA II Programme that has received funding from the EU's H2020 research and innovation programme under the GA No 101017733.

\begin{center}
\textbf{Declarations}
\end{center}
Lukas Cvitkovich is an inventor on a patent application related to this work (Application number: PCT/EP2026/053795).

\appendix

\section{Resonant fraction}
\label{app:E0}

We discuss the normalization factors omitted in Eq.~\eqref{eq:VSisFT} and clarify the meaning of the parameter $\VGe$ (see also App.~B of~\cite{Salamone_kp_2026}).

Revisiting the derivations, we now keep the Bloch parts in Eq.~\eqref{eq:Delta}, giving Eq.~\eqref{eq:VSisFT} in an exact form,
\be
\label{eq:VSisFT2}
\VS = 2 \left|  \int  \diff z |\PsiEnv (z)|^2 u_+^*(z) u_-(z) \QWPot(z) e^{-i 2k_0 z} \right|.
\ee
We assume that the envelope wave function changes slowly over atomic scale and is normalized such that $\sum_n |\PsiEnv(z_n)|^2=1$, whereas the Bloch functions are normalized such that the integral of $|u_\pm|^2(z)$ over a volume corresponding to a single atom is one.
We again insert Eq.~\eqref{eq:discretizedQWPot} into Eq.~\eqref{eq:VSisFT2}, but this time, instead of Dirac delta functions, we consider functions with a finite spatial extent, denoting them $\VGe(z)$. Assuming that such a function is localized on a scale short compared to the variation of the envelope function, we get
\begin{subequations}
\be
\VS = \left|  \sum_m v_0 |\PsiEnv (z_m)|^2 e^{-i 2k_0 z_m} \chi_m \right|,
\ee
where we have defined
\be
\label{eq:chim}
\chi_m =\frac{2}{v_0} \int  \diff z\, \VGe(z-z_m) u_+^*(z) u_-(z) e^{-i 2k_0 (z-z_m)}.
\ee
\end{subequations}
If the functions $u_\pm$ were periodic upon a monolayer distance shift, the energy $\chi_m$ would be independent of the index $m$. This is not the case, since the period of the Bloch functions $u_\pm$ is four times the monolayer width. However, when considering a three-dimensional lattice where Si atoms of monolayer planes are being replaced by Ge randomly, this randomness amounts to averaging of $\chi_m$ over (possibly non-equivalent) atomic positions. The valley splitting is thus proportional to the energy
\be
\label{eq:chimAverage}
E_0 = \left| \overline{\chi_m} \right|,
\ee
where the overline denotes averaging over atoms in the Si crystal.\footnote{In Ref.~\cite{Salamone_kp_2026}, the quantity equivalent to $\chi_m/2$ is denoted by $V_\mathrm{inter}^\mathrm{Ge}$, see their Eq.~(B2) and the authors obtain $V_\mathrm{inter}^\mathrm{Ge}=514$ meV in Si$_{0.7}$Ge$_{0.3}$.}

The evaluation of the integral in Eq.~\eqref{eq:chim} is precluded by the fact that the precise form of $\VGe(z)$ is unknown. It is not simply the Hartree potential of a Ge atom. First, only the difference between such a potential corresponding to a Ge atom and that corresponding to a Si atom should enter Eq.~\eqref{eq:chim}. 
Second, the Hartree potential does not include the non-local terms from exact Hartree-Fock exchange (which are considered in the DFT calculations).
Third, we speculate that the `correct' $\VGe(z)$ is a difference to some hypothetical potential of a virtual crystal Si$_{1-x}$Ge$_x$, with $x$ the fraction of Ge atoms in the vicinity of the site $m$.\footnote{One way to see this is to interpret the results presented in App.~\ref{sec:nonlinearity} as that, for a fixed doping pattern, increasing the Ge content becomes progressively less effective in inducing the valley splitting compared to the prediction of Eq.~\eqref{eq:VSisM}.} 

Due to the uncertainties in the definition of $\VGe(z)$, as well as the difficulties of evaluating such a definition numerically, we resort to extracting $E_0$ from a fit to the tight-binding data as explained in the next Appendix.

\section{Fitting $E_0$}
\label{app:fitting}

\begin{figure}
	\centerline{\includegraphics[width=\linewidth]{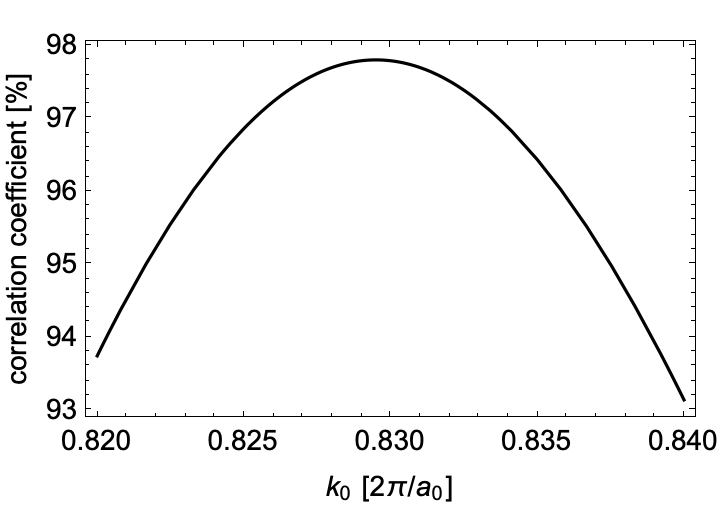}}
    \caption{The correlation coefficient between the valley splitting energy $\VS$ (data plotted in yellow in Fig.~\ref{fig:survey}) and the {\MEFFname}  $\MEFFsymbol(k_0)$, defined in Eq.~\eqref{eq:VSisM}, as a function of $k_0$ used in its definition. 
   The curve attains its maximum at $k_0=0.82928$.%
}
    \label{fig:fittingK0}
\end{figure}

Here we quantitatively assess how closely the tight-binding numerics follow the simple formula in Eq.~\eqref{eq:VSisM}, a linear relation between the valley splitting $\VS$ and the {\MEFFname} $\MEFFsymbol$.  Along the way, we also estimate the proportionality coefficient $E_0$. For simplicity, we use only the random configurations (marked in yellow in Fig.~\ref{fig:survey}) in all fits in this and the next section. 

In a quantum well with many monolayers, a typical situation, the quantity $\MEFFsymbol(k_0)$ is sensitive to the exact value of the momentum $k_0$ of the conduction band minimum used in its definition. 
To illustrate this sensitivity, we plot the correlation coefficient $r$ between $\VS$ and $\MEFFsymbol(k_0)$ as a function of $k_0$ in Fig.~\ref{fig:fittingK0}. It shows a clear maximum that can be located with high accuracy (3 significant digits; see Tab.~\ref{tab:fittedParameters}). The figure confirms the sensitivity to $k_0$: Using a value differing as little as 0.01 from the actual $k_0$ significantly decreases the correlation coefficient $r$, boosting $1-r$ by a factor of three.\footnote{This sensitivity is strong enough to detect subtle structural effects: fitting our model with an added parameter for Ge-induced monolayer widening gives $1.1 \pm 0.3$ pm per 10\%-Ge-doped monolayer, in good agreement with the $0.93$ pm read off directly from the VFF-relaxed atomic geometry in the tight-binding data.}

\begin{figure}
	\centerline{\includegraphics[width=\linewidth]{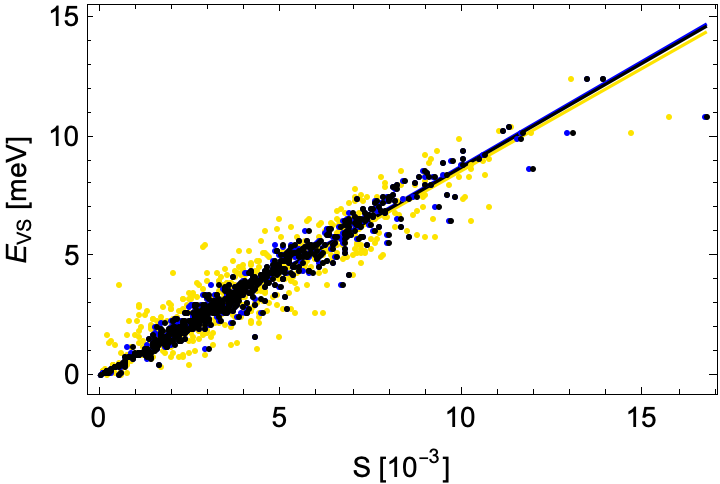}}
    \caption{
    The valley splittings from numerics (only the random configurations, plotted in yellow in Fig.~\ref{fig:survey}) as a function of the {\MEFFname} $\MEFFsymbol$. For the yellow data set, we evaluate $\MEFFsymbol$ assuming $k_0 = 0.84$ as in Fig.~\ref{fig:survey}, for the blue data set, we use $k_0 = 0.8295$. For the black data, we use Eq.~\eqref{eq:VSisMrefined} with $k_0=0.8295$ and $E_0=263$ meV. The lines show the least-squares fits in corresponding colors, with fitted slopes of 860, 880, and 874 (in meV), and one minus the correlation coefficients being 6.77\%, 2.054\%, and 2.037\% for yellow, blue, and black, respectively.%
    }
    \label{fig:fittingE0}
\end{figure}

We replot the random-configurations data from Fig.~\ref{fig:survey} in Fig.~\ref{fig:fittingE0}, for $k_0$ equal to the original choice of $0.84$ (yellow) and the value $0.8295$ from Tab.~\ref{tab:fittedParameters}. The least-squares linear fit for the blue data gives $E_0 \approx 0.9$ eV. Additional data, plotted in black, represent a refined theory for fitting of $k_0$ and $E_0$, and are explained in Appendix \ref{app:refinement}.

\section{Fitting model parameters}
\label{app:fittedParameters}

In the previous section, we looked at the correlation coefficient between the valley splitting observed in tight-binding numerics and predicted by Eq.~\eqref{eq:VSisM} as a function of $k_0$ used in that equation. The result in Fig.~\ref{fig:fittingK0} allows us to fit the value $k_0$. We now extend this analysis in two ways: we use a simple Monte Carlo procedure to obtain the error bars on the best-fit value, and we analyze additional model parameters beyond $k_0$. 

For the former, we perform several (typically 100--1000) minimizations with respect to a given parameter, using half of the data, selected randomly and independently for each run. This gives a set of best-fit values, whose variance we define as the error bar. We also investigate structures with different strains (which can be achieved by varying the Ge concentration in the barrier), for reasons explained in App.~\ref{app:strain}. The results are gathered in Tab.~\ref{tab:fittedParameters}. 

For example, for the 30\% Ge concentration in the barrier, we obtain $k_0 = 0.82953 \pm 0.00030$. The error bar on the estimate obtained in this way should be interpreted as the typical spread of values of the parameter obtained from minimizations for a given amount of VS data. The maximum located through minimization over all data, given in Fig.~\ref{fig:fittingK0}, is thus about one standard deviation from the value found in Tab.~\ref{tab:fittedParameters}, confirming both the value and its error estimate. 

Alternatively to assigning a parameter a best-fit value, 
the presented analysis can be understood as a quantification of the sensitivity of the valley splitting to a model parameter in question. We now comment on the remaining entries in Tab.~\ref{tab:fittedParameters} from this point of view.

We first look at the electric field $E$ and mass $m_l$. One expects a sizable sensitivity of the valence splitting to these parameters, as they define the weights $w_n$ within the quantum well, which enter Eq.~\eqref{eq:VSisM} in a profound way. We confirm this expectation by finding that the best-fit values are close to the tight-binding input of 5 mV/nm for $E$ and the textbook value for $m_l=0.91 m_e$ for the longitudinal mass.\footnote{We reiterate that $m_l$ stands for a parameter of the valley-splitting model expressed by Eq.~\eqref{eq:VSisM}, fitted against the large dataset of valley splittings obtained from tight binding. It does not necessarily reproduce exactly the value that could be extracted directly from the tight-binding band structure, which is a different fitting procedure. For bulk Si, our tight-binding code reproduces the band structure with the textbook value $m_l=0.91\, m_e$.}

Next, we take $\Delta_\mathrm{CB}$, the conduction band offset between the quantum well and the barrier. For 30\% Ge barrier concentration, the expected value is about 150 meV. The minimization gives similar values. On the other hand, considering different Ge concentrations, one expects a linear dependence of $\Delta_\mathrm{CB}$ on it, which is only partially present in Tab.~\ref{tab:fittedParameters}. The valley splitting sensitivity to $\Delta_\mathrm{CB}$ being not large enough to see this trend clearly is not unexpected~\cite{Salamone_kp_2026}, as the conduction band offset enters only through defining the wave function tail within the barrier. 

An alternative probe of the role of this tail concerns the parameter $w_\mathrm{Int}$, defined as follows. We split the contribution of doping layers to the {\MEFFname} into the quantum well and the interface, the former defined as from the $N$ middle monolayers (see Sec.~\ref{sec:QW nomenclature} for the monolayer indexing),
\be
\label{eq:wInt}
\MEFFsymbol = {\MEFFsymbol}_\mathrm{qw} + w_\mathrm{Int} {\MEFFsymbol}_\mathrm{Int}.
\ee
While in reality the Ge scatterers contribute to the {\MEFFname} throughout the heterostructure, including the interface,  corresponding to $w_\mathrm{Int}=1$, the fit reveals whether its influence is statistically detectable. For our quantum well design (with 6 monolayers linearly interpolating between the 30\% Ge concentration in the barrier and 0\% in the quantum well), the interface contribution is ${\MEFFsymbol}_\mathrm{Int} \approx 0.11 \times 10^{-3}$. For valley splittings on the order of meV, which translates to $\MEFFsymbol$ on the order of units times $10^{-3}$, the interface contribution is thus negligible. Nevertheless, we find that it is statistically detectable, since the fitted $w_\mathrm{Int}$ is close to one.

\begin{table}
 \begin{center}
\begin{tabular}{c@{\quad}c@{\quad}c@{\quad}c}
\toprule
strain in QW& 1.35\% &1.15\% & 0.95\%\\
(eq. $\rho_\mathrm{Ge})$  & \multirow{ 1}{*}{35\%} & \multirow{ 1}{*}{30\%} & \multirow{ 1}{*}{25\%}\\
\midrule
$k_0 [2\pi/a_\mathrm{Si}]$ & 0.826(51$\pm$24) & 0.829(53$\pm$30)& 0.831(92$\pm$27)\\ 
$E [\mathrm{V/\mu m}]$ &5.(00$\pm$09) & 5.(14$\pm$10)& 5.(15$\pm$10)\\
$m_l [m_e]$ &0.7(91$\pm$21) & 0.8(31$\pm$25)& 0.8(35$\pm$22)\\
$\Delta_\mathrm{CB} [\mathrm{meV}] $ &167$\pm$8 & 152$\pm$10& 155$\pm$10\\
$w_\mathrm{Int} [1]$ &1.(11$\pm$19) & 1.(09$\pm$17)& 1.(25$\pm$18)\\
\midrule
$E_0 [\mathrm{meV}]$  &  \multirow{ 2}{*}{899$\pm$5} & \multirow{ 2}{*}{891$\pm$7}& \multirow{ 2}{*}{890$\pm$6}\\
from Eq.~\eqref{eq:VSisM} \\\midrule
$E_0 [\mathrm{meV}]$&  \multirow{ 2}{*}{38$\pm$71} &  \multirow{ 2}{*}{263$\pm$195} &  \multirow{ 2}{*}{246$\pm$127}\\
from Eq.~\eqref{eq:VSisMrefined}\\
\bottomrule
\end{tabular}
\end{center}
\caption{
\label{tab:fittedParameters}
\textbf{Fitting model parameters.} The table collects results of model parameter fits for three datasets, differing by the strain (the three strains correspond to equivalent concentrations of Ge in the quantum well barrier being 35, 30, and 25\%, respectively). The model parameters include $k_0$, which enters Eq.~\eqref{eq:VSisM} directly, the three parameters, $E$, $m_l$, and $\Delta_\mathrm{CB}$, which enter Eq.~\eqref{eq:VSisM} indirectly, by defining the weights $w_n$, the parameter $w_\mathrm{Int}$, defined in Eq.~\eqref{eq:wInt}, and $E_0$, obtained from a linear fit using Eq.~\eqref{eq:VSisM} and from a non-linear fit using Eq.~\eqref{eq:VSisMrefined}. The confidence intervals are assigned based on the variance across independent minimization runs (see the text). %
}
\end{table}

\section{A simple theory refinement}

\label{app:refinement}

The variational picture presented in the main text Sec.~\ref{sec:variational} suggests the following refinement of the theory relating the expected valley splitting to the positions of Ge scatterers. Namely, the only variational parameter that we have considered in Sec.~\ref{sec:variational} was the scattering phase $\phi$. We now consider an additional parameter, being the crystal momentum $k$ of the wavefunctions $\psi_\pm$. Previously, we fixed $k=k_0$; now we allow deviations. These deviations come with an energy cost: the kinetic energy at the conduction-band minimum, set by the longitudinal mass, which we take to be $m_l = 0.8 m_e$, with $m_e$ the free-electron mass. The minimization over the phase $\phi$ leads to the same result as before, namely Eq.~\eqref{eq:variationalResult}, and in turn Eq.~\eqref{eq:VSisM}. Adding the energy cost for the new variational degree of freedom then gives the final expression\footnote{We insert the factor of 2 in the second term in the bracket since the quantity $\VS$ is assigned to a pair of states, not a single state.}
\begin{align} \label{eq:VSisMrefined}
\VS = \max_{k} \left(E_0 \left| \sum_{n} \rho_n w_n e^{-2ik z_n} \right| - 2\times\frac{\hbar^2}{2m_l}(k-k_0)^2 \right).
\end{align}
The theory remains simple: a one-dimensional minimization task for each configuration of Ge scatterers.

\begin{figure}
	\centerline{\includegraphics[width=\linewidth]{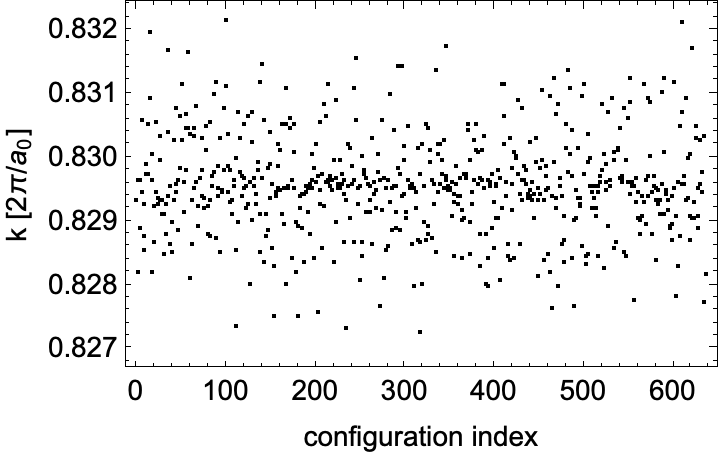}}
    \caption{
    The values of momenta $k$ that maximize the variational energy given in Eq.~\eqref{eq:VSisMrefined}. The x axis is a simple configuration label without special significance. 
        }
    \label{fig:adjustedK}
\end{figure}

We reprocess the tight-binding data from Fig.~\ref{fig:survey} (again, only the random configurations), assigning, for each point individually using a numerical search, its optimal $k$ with which the corresponding fine-tuned {\MEFFname} $\MEFFsymbol(k)$ is evaluated. While we find that we can not increase the correlation coefficient further by this method by a statistically significant amount, we still include the refined {\MEFFname}s in Fig.~\ref{fig:fittingE0} and plot the values of the assigned momenta $k$ in Fig.~\ref{fig:adjustedK} as an illustration.

Having Eq.~\eqref{eq:VSisMrefined} allows us to extend the correlation-maximization analysis to $E_0$ as a model parameter. We fit $E_0$ using Eq.~\eqref{eq:VSisMrefined} in the way described in the previous section. The results are included in Tab.~\ref{tab:fittedParameters}. While we obtain values in the correct order of magnitude, they are still a factor of $O(1)$ smaller than the values for $E_0$ that one can infer from a least-squares fit to the data in Fig.~\ref{fig:fittingE0}. 

We thus conclude that Eq.~\eqref{eq:VSisMrefined} is not an improvement that would increase the theory prediction power.  We believe that one of the reasons is that our doping concentration is relatively high (10\%),\footnote{The better alignment of the effective-mass model and the tight-binding data presented in Fig.~1f of Ref.~\cite{losert_practical_2023} was observed for an order of magnitude lower doping concentrations.}
invalidating the simple perturbative picture for the valley splitting behind Eq.~\eqref{eq:VSisM}. Our DFT data, such as shown in Fig.~\ref{fig:DFT}, support this speculation.

Nevertheless, our main conclusion is that, supported by the cross-checks with respect to the model parameters done in Apps.~\ref{app:fittedParameters} and \ref{app:refinement}, the correlation coefficients close to 100\% observed over large data sets mean that, regarding trends and estimates, the microscopic numerics (tight binding and DFT) are entirely disposable. One can predict the valley splitting \textit{statistics} from the simple formula for the quantum well {\MEFFname} given by Eq.~\eqref{eq:VSisM} with good accuracy.

\section{Monolayer step influence}
\label{app:step}

Monoatomic steps in the quantum well layers were proposed long ago as a possible explanation for unexpectedly small valley splittings observed experimentally \cite{Ando_valley_1979}. Since then, the role of monoatomic steps has been discussed at length, and they are believed to be a possible issue for valley splitting, potentially causing strong suppression. 

Our numerics implements quantum wells with statistically uniform properties in lateral dimensions, and thus we can not investigate monoatomic steps directly. Nevertheless, due to their potential danger, we aim to provide at least rough estimates. 

Let us first explain the physics behind our approach. We again use a variational approach. The system ground state corresponds to the minimum (over the variational parameters) of the energy, which means the maximum of $\VS$, since the ground state energy equals $\mathrm{const}-\VS/2$. If the electron sits directly on the step, $\VS$ is suppressed. The electron can increase it, and thus lower its energy, by moving laterally off the step. Thus, there is a force pushing the electron away from the step. The system ground state is at the position where this repulsive force due to the step equals the opposite force exerted by the confinement potential. In the following, we evaluate these two competing contributions to the electron's total energy.

\begin{figure}
	\centerline{\includegraphics[width=0.7\linewidth]{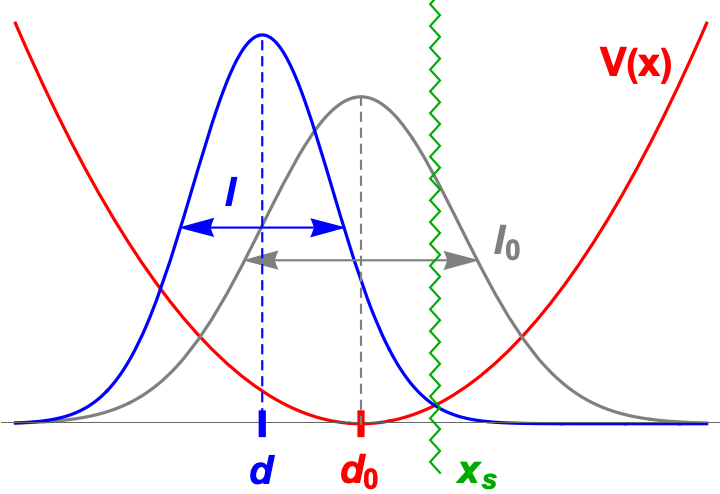}}
    \caption{Schematic of a model to estimate the influence of an atomic step. The confining quadratic potential centered at $d_0$ is in red. The corresponding ground-state wave function is in gray, centered at $d_0$, and has width $l_0$. The variational wave function has the same (Gaussian) shape, but its center is shifted to $d$, away from the step located at $x_s$, and squeezed to $l$.
       }
    \label{fig:schematic}
\end{figure}

We first extend the variational basis functions $\psi_\pm$ by the in-plane coordinate $x$. Figure \ref{fig:schematic} illustrates the basic elements of the model. We assume that along the $x$ coordinate, the electron is confined by a quadratic confinement $V(x)=(\hbar^2/2m_t l_0^4) (x-d_0)^2$ with $l_0$ the confinement length, $d_0$ the dot center, and $m_t$ the transverse mass. With the kinetic energy $T(x) = - (\hbar^2/2m_t) \partial_x^2$, the ground state wave function is
\be
\label{eq:inplaneWF}
\Psi(x) = \frac{1}{\pi^{1/4} l_0^{1/2}}\exp\left( -\frac{(x-d_0)^2}{2l_0^2} \right).
\ee

We now consider that a monoatomic step running along the y-axis (not explicitly present in the model) is located at $x=x_s$. In the simplest picture, it displaces all $z$ coordinates of Ge scatterers of a given `monolayer' with $x$ coordinate $x>x_s$ by one monolayer distance $a_\mathrm{Si}/4$ with respect to those with $x<x_s$. For the wave function given in Eq.~\eqref{eq:inplaneWF}, this shift leads to a strong reduction (up to above 70\%, see below) of the {\MEFFname} $\MEFFsymbol$ if $x_s$ is close to the dot center, since the factor $\exp(i k_0 a_\mathrm{Si}/2)$ is close to minus one.

\begin{figure}
	\centerline{\includegraphics[width=0.9\linewidth]{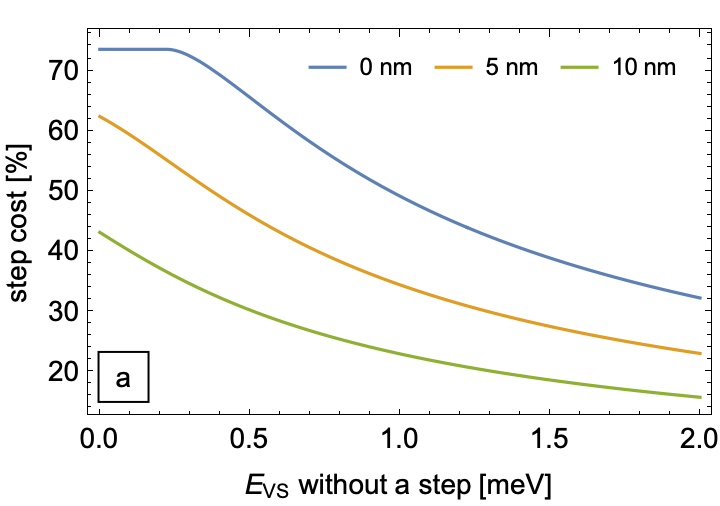}}
	\centerline{\includegraphics[width=0.9\linewidth]{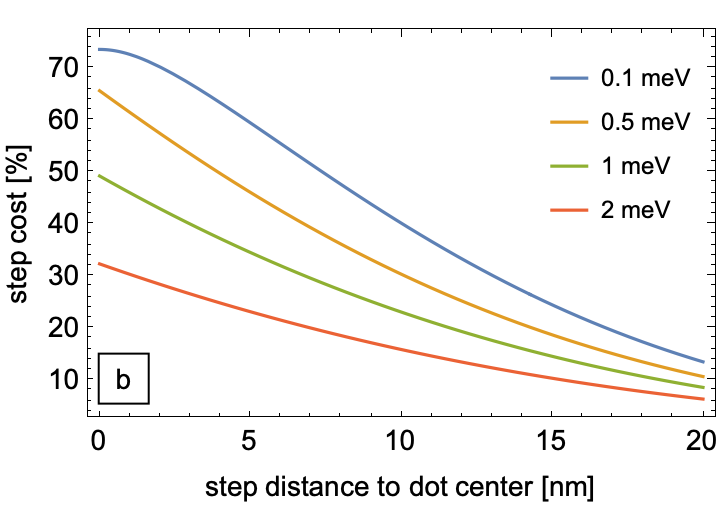}}
    \caption{The fractional cost of an atomic step in the quantum well. The cost (y axis) shows $1-\VS^\mathrm{step}/\VS^\mathrm{bare}$, with $\VS^\mathrm{bare}$ being a fixed parameter and $\VS^\mathrm{step}$ is calculated according to Eq.~\eqref{eq:VSisMrefined2}. 
    In a) $\VS^\mathrm{bare}$ is the x-axis variable, and the three colored curves correspond to different step locations, with the value for  $x_s - d_0$ as given in the legend. 
    In b) $x_s - d_0$ is the x-axis variable, and $\VS^\mathrm{bare}$ is fixed for each colored curve according to the legend.
    We use $m_t=0.19\, m_e$ and $l_0=20$ nm, corresponding to an in-plane orbital excitation energy $\hbar^2/m_tl_0^2 = 1$ meV.
       }
    \label{fig:stepCost}
\end{figure}

We now adopt Eq.~\eqref{eq:inplaneWF} as the in-plane wave function profile and promote parameters $d_0$ and $l_0$ to variational parameters, dropping the subscript '0' in that case. Figure \ref{fig:schematic} shows a simple way of increasing the variational energy gain by, effectively, decreasing the effect of the step. The wave function can be displaced away from the step (expressed by the parameter $d \neq d_0$), and squeezed (expressed by the parameter $l\neq l_0$). Both of these changes increase the variational energy gain by increasing the {\MEFFname} $\MEFFsymbol(d,l)$, and come with an energy cost $\delta E(d,l)$. We get the analog of Eq.~\eqref{eq:VSisMrefined}
\begin{align} \label{eq:VSisMrefined2}
\VS = \max_{d,l} \Big( E_0 \MEFFsymbol(d,l) - 2 \times \delta E(d,l) \Big).
\end{align}
We define the energy cost as the increase in the expectation value of the Hamiltonian for the variational wave function, $\delta E(d,l) \approx \langle \Psi | T + V | \Psi \rangle$. Simple algebra gives 
\begin{subequations} \label{eq:dlEquations}
\be
 \delta E(d,l) =
 \frac{1}{4} \frac{\hbar^2}{m_t l_0^2} \left( \frac{l_0^2}{l^2} + \frac{l^2}{l_0^2} - 2 + 2 \frac{(d-d_0)^2}{l_0^2} \right).
\ee
For the {\MEFFname}, we get
\be
E_0 \MEFFsymbol(d,l) = \VS^\mathrm{bare} \left| w_{>x_s} + (1-w_{>x_s}) e^{ik_0 a_\mathrm{Si}/2} \right|,
\ee
where $\VS^\mathrm{bare}$ is the valley splitting without any step, for which we use Eq.~\eqref{eq:VSisM},
and $w_{>x_s}$ is the weight of the wave function beyond the step
\be
w_{>x_s} = \int_{x_s}^\infty \mathrm{d} x |\Psi(x)|^2 = \frac{1}{2}\left(1+ \mathrm{Erf}\left[\frac{d-x_s}{l} \right] \right).
\ee
\end{subequations}
Equations \eqref{eq:dlEquations} define Eq.~\eqref{eq:VSisMrefined2} as a simple minimization problem. The resulting variational energy depends only on the bare (that is, no step) valley splitting, the in-plane confinement length and mass, and the distance of the atomic step from the quantum-dot center. In particular, it does not depend on the details of the Ge configuration, what allows for a concise presentation of the main features.

We depict the reduction of the valley splitting due to an atomic step in Fig.~\ref{fig:stepCost}. Panels a and b show, respectively, the reduction as a function of the bare valley splitting and the step position, with the other parameter fixed to a few typical values. A step through the dot center can substantially suppress the valley splitting when it is small (below a few hundred $\mu$eV), a well-known conclusion. However, once the bare valley splitting reaches the meV scale, the reduction from a step is a much smaller fraction. With $\VS$ of 1 meV, a maximal reduction is 50\%, taking place if the step runs exactly through the dot center. For larger bare splittings, the reduction is even less important. We thus conclude that once the bare splitting gets into the meV range, the atomic steps in the quantum well do not pose a serious threat.

\section{Stability of the magic numbers 5 and 7}

\begin{figure}
	\centerline{\includegraphics[width=\linewidth]{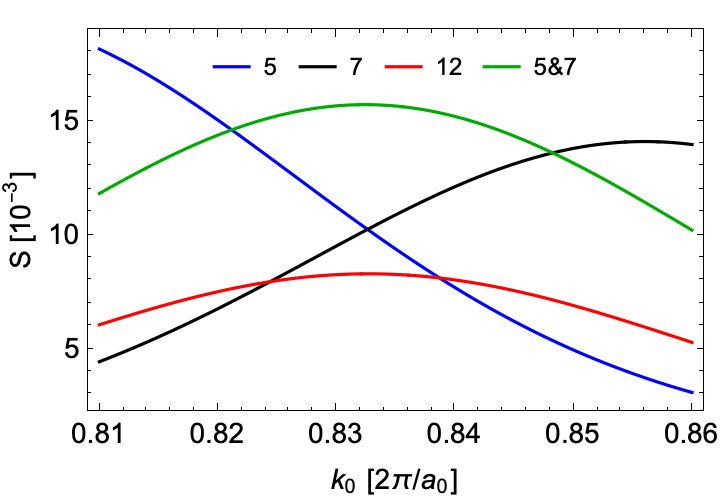}}
    \caption{The {\MEFFname} calculated using Eq.~\eqref{eq:VSisM} as a function of $k_0$ for three periodic profiles with periods $p$ equal to 5, 7, and 12, and a profile alternating 5- and 7-monolayer steps between consecutive Ge scatterers.
    }
    \label{fig:magicStability}
\end{figure}

In the main text, we have shown that a profile composed of alternating 5- and 7-monolayer steps between consecutive Ge scatterers is exceptionally promising. Fortuitously, such a type of profile, namely one based on 12 monolayer blocks, is stable with respect to the actual value of $k_0$ around 0.83. To demonstrate this, in Fig.~\ref{fig:magicStability} we plot the {\MEFFname} as a function of $k_0$ for the profiles with period $p$ (the notation of Fig.~\ref{fig:ideal}) equal to 5, 7, 12, and the alternating 5-7-5-7-5-7 one. The $p=12$ curve reaches a maximum at $k_0 =10/12 \approx 0.833$, since here the product of $k_0$ and $p$ becomes an integer. At its extremum, the {\MEFFname} is thus insensitive to small deviations of $k_0$ from this value. Unlike for periodic profiles with $p=5$ or $p=7$, this stability is inherited by the more complex profile 5-7-5-7-5-7 and its variants (see App.~\ref{app:finalSuggestion}).

\section{Random doping gives $\VS \propto \sqrt{M}$}
\label{app:randomWalk}

\begin{figure}
	\centerline{\includegraphics[width=\linewidth]{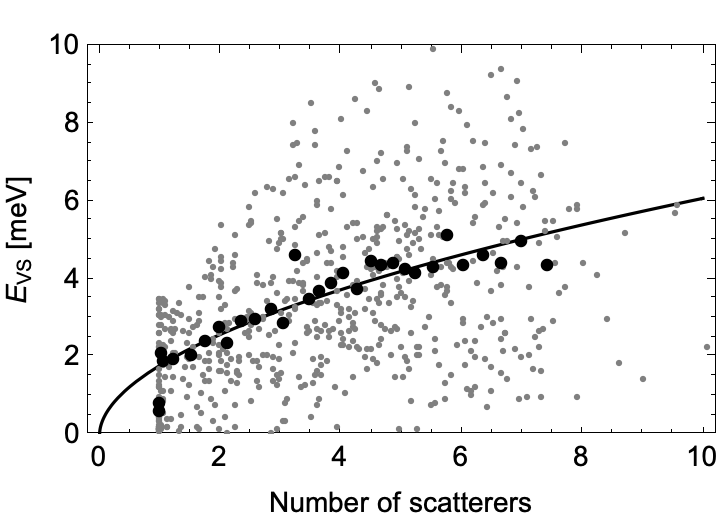}}
    \caption{ The valley splittings from numerics (only the random configurations, plotted in yellow in Fig.~\ref{fig:survey}) as a function of the effective number of Ge scatterers defined in Eq.~\eqref{eq:MeffDefinition}. Each gray point corresponds to one of the yellow points from Fig.~\ref{fig:survey}. Each black point stands for a set of 20 gray points, placed at the coordinates corresponding to the center of mass of the set. The line is a non-linear fit to a power law $VS = c M_\mathrm{eff}^p$ with $c$ and $p$ the fit parameters. The fit gave $c=1.8 \pm 0.12$ and $p=0.53\pm0.04$.
 }
    \label{fig:randomWalk}
\end{figure}

The resonant scattering picture suggests that without any optimization, the total accumulated complex backscattering amplitude is a result of a random walk and thus will vary as $\MEFFsymbol \propto \sqrt{M}$, with $M$ the total number of Ge Scatterers. To examine this prediction, we replot the data of Fig.~\ref{fig:survey} (again, only the random configurations plotted in yellow) as a function of $M_\mathrm{eff}$, the effective number of Ge scatterers. Instead of simply being the number $M$, we use $M_\mathrm{eff}$ to reflect the fact that the wave function is not uniform throughout the quantum well. We use the definition based on the `inverse participation ratio' \footnote {This definition leads to artifacts for configurations with a single Ge plane, which are all assigned $M_\mathrm{eff}=1$, irrespective of the actual wave function value. Since this issue does not prevent us from identifying the square-root dependence, we ignore it.}
\be
\label{eq:MeffDefinition}
M_\mathrm{eff} = \frac{(\sum_n w_n \rho_n)^2}{\sum_n (w_n \rho_n)^2}.
\ee
The resulting data are plotted in Fig.~\ref{fig:randomWalk} and show a large spread. To extract the trend, we fit them to the formula $c M_\mathrm{eff}^p$. We get $p=0.53\pm 0.04$, confirming the expected square root dependence on average. As a more visual illustration, we sort the data plotted in gray according to $M_\mathrm{eff}$, merge 20 consecutive values into a set, and plot the arithmetic average of both $M_\mathrm{eff}$ and $\VS$ for each set as a black thick point. These points clearly show a square-root dependence.

One could follow the random-walk idea further and obtain a prediction for the valley splitting in a uniformly doped quantum well,
\be \begin{split}
\VS &\approx E_0 \sqrt{|\MEFFsymbol |^2}\
= E_0 \sqrt{  \sum_a w_a^2 \mathrm{var} \rho_a },
\end{split} \ee
where the index $a$ now enumerates atoms in the three-dimensional space, covering both out-of-plane and in-plane coordinates. Using $\mathrm{var}\rho_a = \rho(1-\rho)\approx\rho$, with $\rho$ being the mean doping density, and changing the discrete sums to integrals for the wave functions with $v_0=a_\mathrm{Si}^3/8$, the volume per atom, we immediately get
\be \begin{split}
\VS &\approx E_0 \sqrt{ \frac{a_\mathrm{Si}^3}{8} \int \diff z |\Psi_z|^4 \int \diff x \diff y |\Psi_{x,y}|^4 } \,\,\, \sqrt{\rho}.
\end{split} \ee
The integral for the ground state of a triangular confinement created by an electric field $E$ with an associated length given by $l_E^3 = \hbar^2/2m_l e E$ (see Appendix A.1 in Ref.~\cite{stano_orbital_2019}) is
\be
\int \diff z \, |\Psi_z|^4= \frac{0.41}{l_E},
\ee
and for Gaussian in-plane wave functions with confinement lengths $l_x$  and $l_y$, it is
\be
\int \diff x \diff y \, |\Psi_{x,y}|^4 = \frac{1}{2\pi l_x l_y}.
\ee
We finally arrive at
\be
\VS \approx E_0 \sqrt{0.41\frac{a_\mathrm{Si}^3}{16 \pi l_E l_x l_y} }\sqrt{\rho}.
\ee
Inserting our typical parameters, namely $E=5$ mV/nm, $l_x=l_y=20$ nm and $\rho=10\%$, the formula gives valley splitting of 300 $\mu$eV. It agrees to 10-20\% with the dependency given in Fig.~8 in Ref.~\cite{losert_practical_2023} (the pink curve; they use $l_x=l_y$ that are $\sqrt{2}$ smaller than ours).

\begin{figure*}
\includegraphics[width=0.49\linewidth]{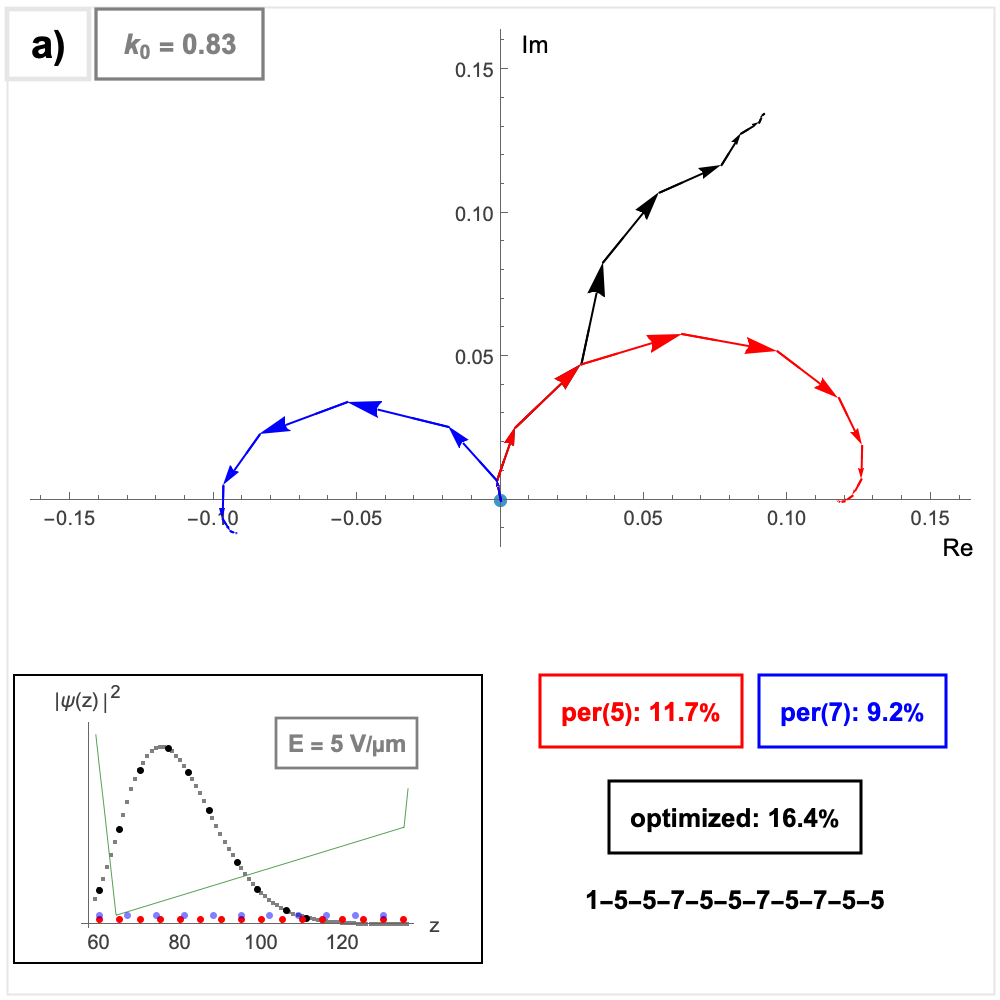}
\includegraphics[width=0.49\linewidth]{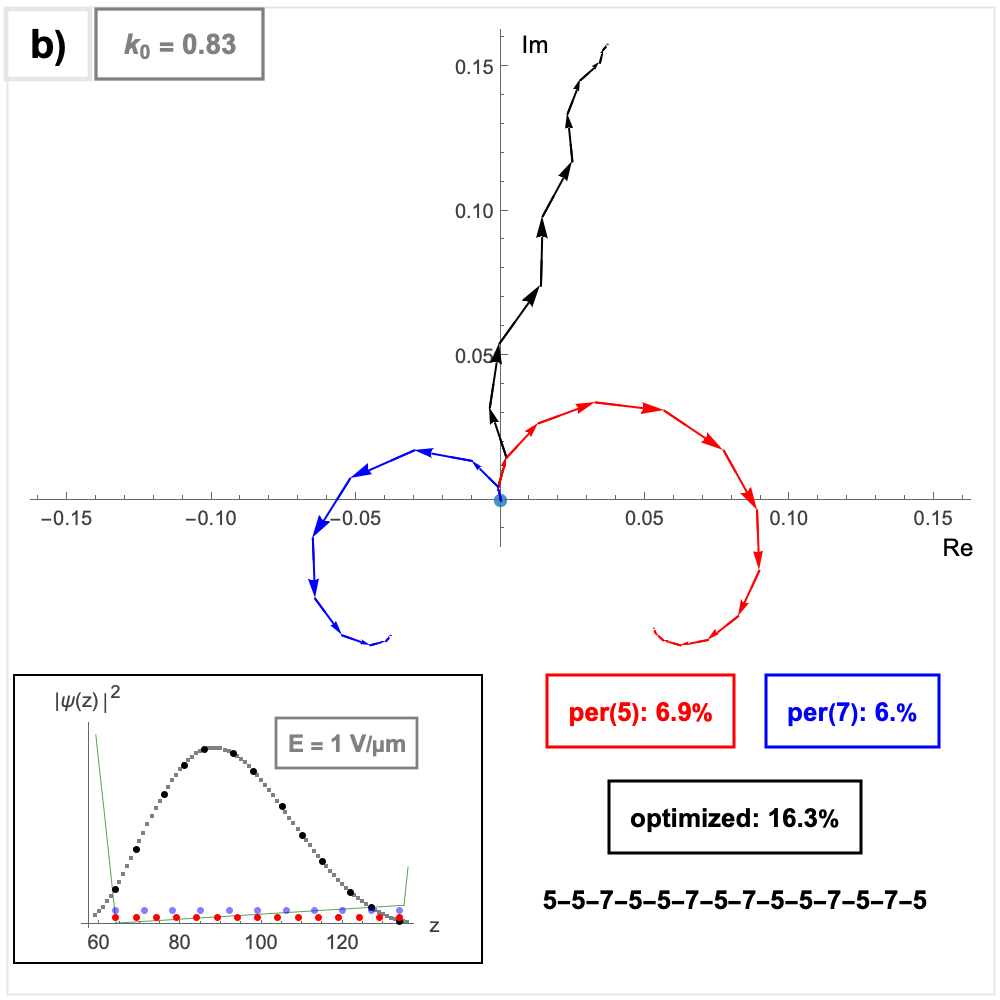}\\
\vspace{0.0cm}
\includegraphics[width=0.49\linewidth]{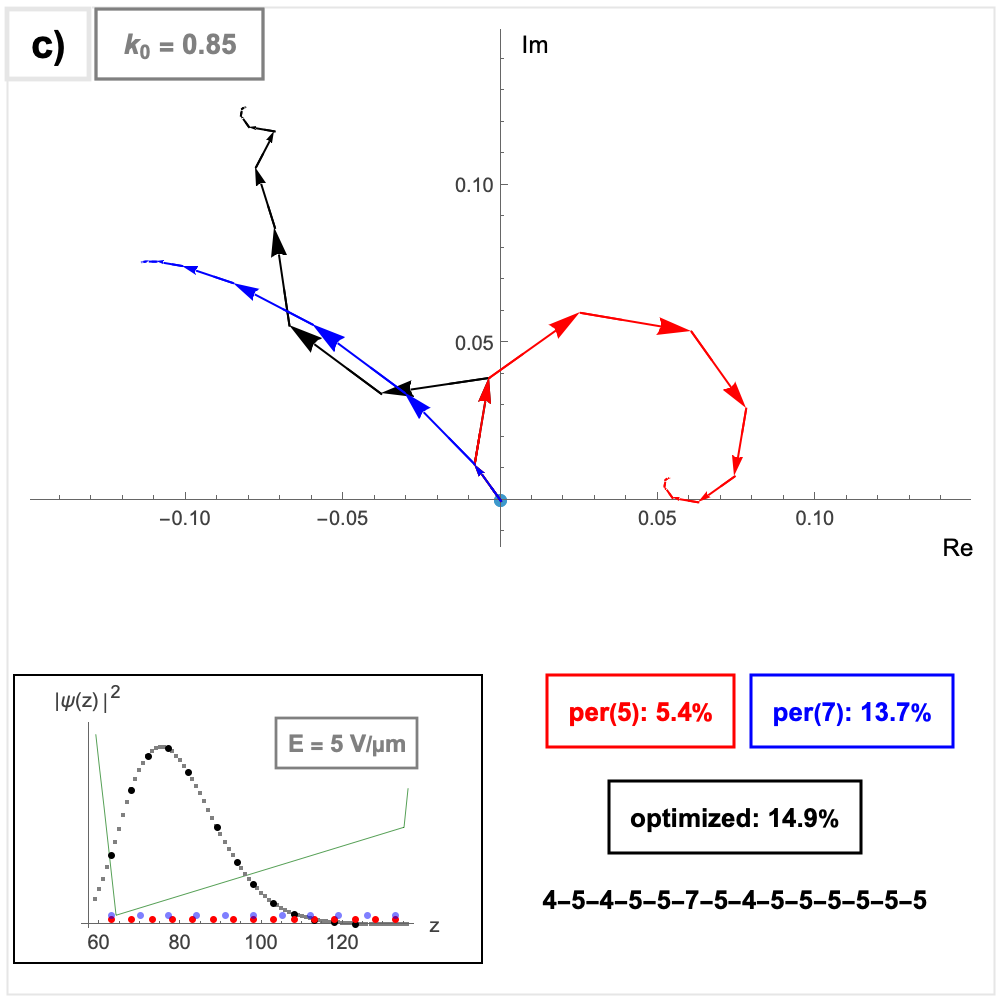}
\includegraphics[width=0.49\linewidth]{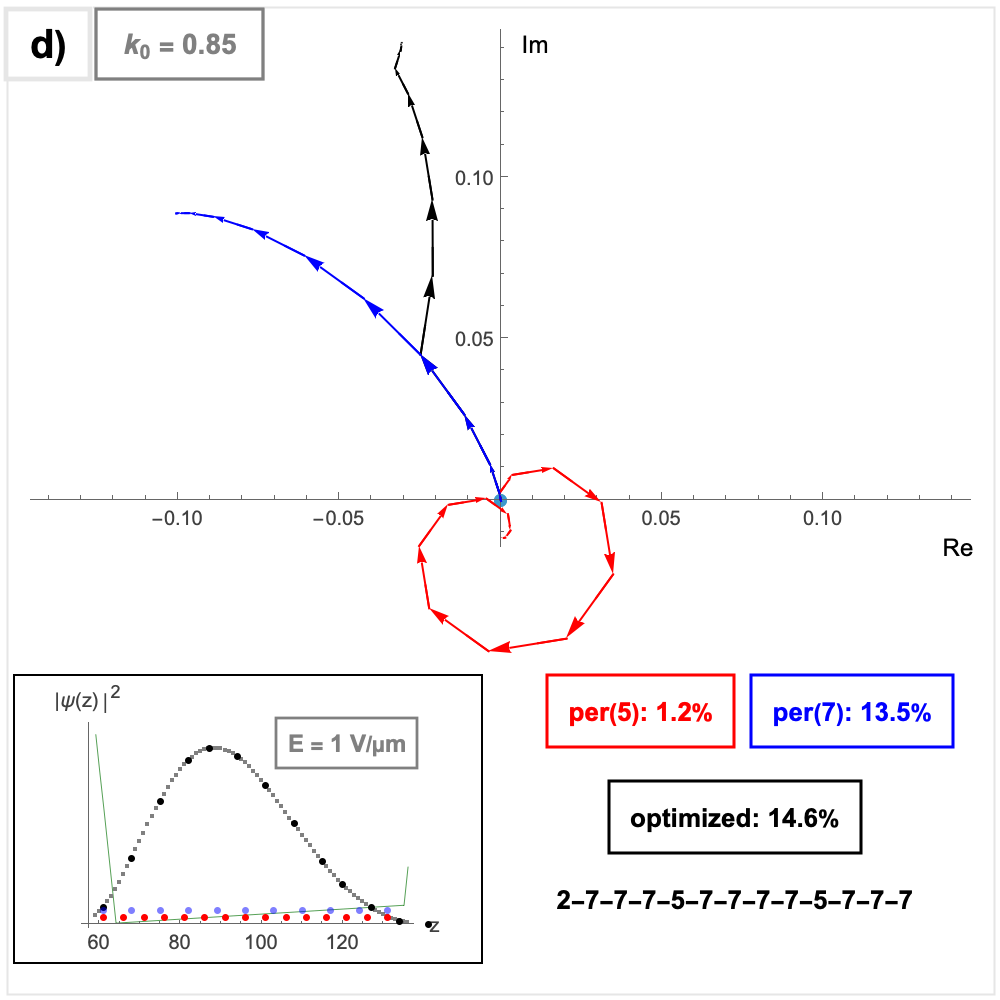}
\caption{\textbf{Visualization.} There are four panels, differing in values for $k_0$ and $E$, as given in framed boxes. Each panel shows the complex plane (axes denoted as Re and Im) with some framed insets in the lower half. The main object of a panel is a path composed of concatenated arrows. The path represents the complex number defining $\MEFFsymbol$ in Eq.~\eqref{eq:VSisM}, and each arrow represents a complex number $\rho_m w_m \exp(-2ik_0 z_m)$ for one $m$. We fix $\rho_m=1$ representing monolayers with (unrealistic) 100\% Ge doping. Three colors represent three different Ge placement patterns. In red and blue, doping monolayers are placed periodically, with periods $p=5$ (red) and $p=7$ (blue). In black, the displacements of consecutive doping-monolayer distances are shown in the bottom-right panel. The lower-left inset shows the wave-function profile (gray points) and the confinement potential (green lines). Each point represents a monolayer site, with sites occupied by Ge denoted by the corresponding colors. The bottom-right framed boxes give, in corresponding color, the resulting $\MEFFsymbol$. We do not consider spacings smaller than 4 monolayers, which can in some cases slightly increase the {\MEFFname} (for example, replacing some of the `7's in the above lists by the pair 2 and 5, and similar).
}
\label{fig:visualization}
\end{figure*}

\section{Visualization, best patterns, and upper bounds on achievable valley splitting}

Here we provide a visualization of the {\MEFFname} $\MEFFsymbol$ defined in Eq.~\eqref{eq:VSisM}. We also give upper bounds on the achievable value of $\MEFFsymbol$. Finally, we provide our best suggestions for configurations that maximize $\MEFFsymbol$.

\subsection{Visualization of Eq.~\eqref{eq:VSisSUM}}

For simplicity, we remain with the one-dimensional model, where the wave function is defined on discrete sites displaced by a monolayer distance $a_\mathrm{Si}/4$, with each site occupied by either Si or Ge. In this scenario, the {\MEFFname} is described by Eq.~\eqref{eq:VSisSUM}. Our goal is to understand how the choices for positions of the Ge sites influence the sum of complex numbers inside the absolute value in Eq.~\eqref{eq:VSisSUM}. 

We consider several combinations of basic parameters of the problem, namely the momentum $k_0$, the wave function shape (dependent on the electric field $E$ and the width of the quantum well Si region), and the patterns of the Ge positions. Each such choice will be presented in a figure (one of four panels of Fig.~\ref{fig:visualization}) with the following content. The main plot of a panel represents a complex plane, where $\MEFFsymbol$ is resolved into the individual contributions, each representing one complex number $w_m \exp(2ik_0 z_m)$ by an arrow. There are three Ge placement patterns distinguished by colors: In red and blue Ge are placed periodically, with the period of $p=5$ (red) and $p=7$ (blue). In black, the positions of Ge scatterers were hand-tuned, with the consecutive distances given at the bottom right. The lower-left inset shows the wave function profile, with the sites occupied by Ge denoted by dots in corresponding colors. The bottom-right framed boxes give, in corresponding colors, the resulting length of the arrow composition, that is, the distance from the start of the first arrow to the end of the last. Thus, these numbers equal $\MEFFsymbol$.\footnote{For the hypothetical case of $\rho_m=1$. More realistic {\MEFFname}s follow upon multiplying by the actual concentration of the doping layers, for example $\rho_m=0.1$ used in Fig.~\ref{fig:survey}.}

With the figure elements described, let us look at the specific cases, starting with Fig.~\ref{fig:visualization}a. It corresponds to the parameters adopted in the main text, namely $k_0=0.83$ and $E=5$ mV/nm. We observe that the consecutive arrows in the periodic patterns rotate relative to one another. This constant rotation arises from the incommensurability of $p$ and $2/k_0$. For red ($p=5$), the rotation is clockwise, since $k_0p/2\approx 2.075$ is slightly larger than an integer, and counterclockwise for blue ($p=7$), where $k_0p/2\approx 2.905$ is slightly smaller than an integer. For paths that are built from many arrows, this rotation accumulates and limits the path length. Since the two rotations are approximately opposite, alternating 5- and 7-monolayer distances yields an approximately straight path with maximal length. We find that breaking even that periodicity allows us to stretch the maximal distance further. The sequence of distances for the configuration that yields the maximal length we found (by trial and error) is given below the black-framed box.

Next, we move to Fig.~\ref{fig:visualization}b, which shows the analogous quantities for a quantum well with a smaller electric field. With that, the wavefunction extends further along $z$, as is appreciable from the left-bottom inset. It is now mostly confined by the Si region width, rather than by the electric field. In this case, the weights $w_n$ fall off more slowly with the index $n$. As a consequence, the paths are composed of more arrows, and the detrimental effects of the incommensurability-induced rotation are more severe. The simple periodic patterns do not reach even 50\% of the value of the optimized path where the rotations are intentionally compensated.

Figure \ref{fig:visualization} panels c and d are analogous to panels a and b, but for the change of the value of $k_0$ to $0.85$. 
This change improves the commensurability for $p=7$ and worsens it for $p=5$. As a consequence, the pattern built on the period $p=5$ becomes particularly ineffective, while the periodic $p=7$ pattern is acceptable. We find that by judicious choices, the resulting $\MEFFsymbol$ can be stretched further, by about 10\% in relative scale. The exact value depends on the wave function profile. For example, for the $E = 5$ mV/nm quantum well, the best pattern we found is highly non-trivial, employing, apart from 5 and 7, also 4 monolayer steps. 

\subsection{The best patterns}
\label{app:optimization}
\label{app:finalSuggestion}

The best patterns for the four parameter sets analyzed in Fig.~\ref{fig:visualization} are shown in the corresponding bottom-right panels as a series of monolayer distances between consecutive Ge monolayers. We copy them to Tab.~\ref{tab:best}. The resulting {\MEFFname} $\MEFFsymbol$ is above 16\% for the $E=5$ mV/nm quantum well and a comparable value of above 15\% for the $E=1$ mV/nm one. These values are close to our estimates of the upper bounds on $\MEFFsymbol$ (see the next subsection), and thus we do not expect that they can be improved much further. We also note that the values plotted in black were found by hand, by trial and error, within a minute each, based on the strategy of `make the black path as straight as possible'. In contrast, we also used brute-force numerical maximization (one implemented in Mathematica and one in Python). However, the execution time of such a numerical search grows exponentially with the number of Ge monolayers considered, and our hardware and software limited us to configurations with up to 8 scatterers (the results plotted in Fig.~\ref{fig:survey} in black). Despite their complexity, these brute-force methods did not find configurations significantly better than the ones plotted in Fig.~\ref{fig:visualization}.%

\subsection{The upper limits on achievable {\MEFFname}} 

We now derive rough upper-limit estimates for the achievable {\MEFFname}. We assume that the quantum well is not very narrow (either due to a narrow Si region or due to a large electric field) so that the wave function extends over many monolayers (a number much larger than 1). In this limit, we estimate the upper limit on $\MEFFsymbol$ by neglecting incommensurability in certain hypothetical patterns, which can only decrease the corresponding $\MEFFsymbol$.

We first ignore the incommensurability altogether and assume that each time the oscillating phase $\exp(i 2 k_0 z)$ attains a given value, say 1, there is, coincidentally, a monolayer, which can be doped. In this case, the fraction of $4/2k_0$ of all monolayers (which have linear density $4/a_\mathrm{Si}$) would be doped, resulting in $\MEFFsymbol=41.5\%$. This value is unrealistic, since $2k_0a_\mathrm{Si}/4$ is far from an integer multiple of $2\pi$. 

To arrive at a better upper bound, we consider the hypothetical value $k_0=0.8$. In this case, the phase accumulated over 5 monolayers is an exact integer, namely 2, in units of $2\pi$. That is, doping every fifth monolayer, all phases in Eq.~\eqref{eq:VSisSUM} are exactly resonant. Since one fifth of the wave function definition sites are covered by Ge, we get $\MEFFsymbol=20\%$. 

The choice $0.8$ is still far off the realistic value expected for $k_0$. Moving to the next case, we consider $k_0=0.8\overline{3}$. Here, the phase accumulates to an exact integer 5 over 12 monolayers. Doping every 12-th monolayer gives $\MEFFsymbol=8.3\%$. One cannot get twice as much since, 5 being an odd integer, the phase factor at the sixth monolayer is -1 and leads to an almost perfect destructive interference (which gives a simple explanation why $p=6$ in Fig.~\ref{fig:ideal} is the smallest value among all plotted ones). However, by combining 5 and 7 monolayer distances, one can approach an effective doping-layer density of 1/6. Such density corresponds to $\MEFFsymbol=16.7\%$. Finally, we consider the next perfect commensurability, achieved at $k_0=0.857\cdots$ with $p=7$. The doping monolayer density 1/7 corresponds to $\MEFFsymbol=14.3\%$.   

We thus arrive at the following estimates for the upper limits on the {\MEFFname}. With $k_0$ close to 0.83, the maximal achievable $\MEFFsymbol$ is close to 16.6\%. With $k_0$ close to 0.85, the maximum is $14.3\%$. We find it remarkable that the values given in Fig.~\ref{fig:visualization}, which we found through the simple trial-and-error maximization, are very close to these upper limits.\footnote{The fact that the values found are slightly larger than the upper-limit estimates is due to not fully respecting the assumption of the wave-function extending over many monolayers. In other words, this is an artifact of having a narrow quantum well. In an extreme case, one could consider a quantum well SiGe/Si/SiGe with a single monolayer within the "Si" region, and then occupying that site with Ge. Formally, we get $\MEFFsymbol=100\%$. However, this has nothing to do with the valley splitting anymore, with the assumption of perturbative effects needed for the relation between $\VS$ and $\MEFFsymbol$ being completely destroyed.} Based on this, we believe that there is little hope in finding patterns with much higher $\MEFFsymbol$. We speculate that the maximal achievable {\MEFFname}, and thus the valley splitting, will be set by the balance of the pattern complexity and its experimental realizability. From Tab.~\ref{tab:best}, we conclude that mastering a controlled separation of consecutive Ge doping peaks by 5 and 7 monolayers would provide a decisive advantage allowing realization of wafers with valley splittings in the meV scale.

\section{Finetuning of valley splitting by strain}
\label{app:strain}

Previously, we mentioned that the {\MEFFname} $\MEFFsymbol{}$ for a given doping profile is sensitive to the value of the wave vector $k_0$ at which the conduction band has a minimum. This sensitivity can be appreciated from Fig.~\ref{fig:magicStability}, where $\MEFFsymbol$ is plotted as a function of $k_0$ around 0.83, or from Fig.~\ref{fig:visualization}, where the doping layers' contribution in assembling the {\MEFFname} is depicted for two different values of $k_0$. Based on this sensitivity, one is led to consider fine-tuning of $k_0$ to a value beneficial to some simple profiles, for example purely periodic ones with period $p$ considered in Fig.~\ref{fig:ideal}. The {\MEFFname} would be maximized by tuning $k_0$ equal to 0.8 if $p=5$, to $0.8\overline{3}$ if $p=12$, to $0.857\cdots$ if $p=7$, and so on. In this section, we examine how much one can change $k_0$ by strain, as the most obvious option to consider.

We therefore analyze the shift in $k_0$ with strain. 
In the above calculations, the silicon crystal is strained by 1.15\%. To assess the influence of small strain variations, we superimpose an additional strain of $\pm 0.2$\,\%\footnote{We choose this value as equivalent to a 5\% change of the concentration of Ge in the barrier. In our simulations, the strain is externally imposed, and the actual Ge concentration in the barrier is fixed at 30\%.} relative to this reference state.
We first perform a symmetry analysis to show that strain does shift $k_0$,
and then estimate the effective magnitude from tight-binding numerics. We exploit the fact that the tight-binding data on valley splitting allow one to fit the value of $k_0$ with high accuracy, as demonstrated in Fig.~\ref{fig:fittingK0} and App.~\ref{app:fittedParameters}. 

We estimate that changing the strain by 0.1\% (equivalently, changing the Ge concentration in the barrier by 2.5\%) shifts $k_0$ by -0.00125. How beneficial the $k_0$ tuning is will ultimately depend on the actual value of $k_0$ in silicon, the achievable doping patterns, and the extent to which the barrier Ge concentration can be tuned.

\subsection{Symmetry analysis}

We perform a simplified procedure, ignoring that the diamond lattice is nonsymmorphic and adopting its factor group $O_h$ as the system symmetry group. Accordingly, we investigate what Hamiltonian terms are allowed assuming $O_h$ or its subgroups. Concerning the Hamiltonian, we are interested in terms that shift the conduction-band minimum. The dispersion around the minima can be written as\footnote{Here, `low' stands for a `low-energy model'.} 
\be
\label{eq:Hlow}
H_\mathrm{low,\pm} = \frac{\hbar^2}{2m_l} (\mathbf{k} \mp \mathbf{k}_0)^2,
\ee
where the sign subscript discriminates the two minima and $\mathbf{k}_0$ is along $z$. Expanding the square, one sees that a shift of the dispersion curve minimum arises upon adding to the Hamiltonian a new term of the form $\mathbf{k} \cdot \mathbf{v}_0$, where $\mathbf{v}_0$ is a fixed vector. Alternatively, one can introduce a vector of Pauli matrices $\{\tau_x, \tau_y, \tau_z\}$ representing pseudo-spin operators acting on the valley degree of freedom. It differs from spin in that the time-reversal symmetry, as well as inversion symmetry, does interchange the two valleys (both operations sending $k_z$ to $-k_z$) and thus the triple $\boldsymbol{\tau}$ is a polar vector and not an axial pseudo-vector. With this setting, we ask whether one can generate
\be
\delta H_\mathrm{low} \propto c \tau_z k_z, 
\ee
by strain, that is, with the prefactor $c$ proportional to components of the strain tensor $\epsilon_{ij}$.

To check the soundness of the procedure, we first look at several obvious results. First, the term present in the silicon dispersion Hamiltonian Eq.~\eqref{eq:Hlow} is indeed allowed,
\be
\delta H_\mathrm{low} \propto \boldsymbol{\tau} \cdot \mathbf{k} \qquad \mathrm{is\, allowed\,in\,O_h}.
\ee
Second, due to inversion symmetry, the strain should not shift the two minima uniformly. Indeed, for any $i,j$,
\be
\delta H_\mathrm{low} \propto \tau_0 k_z \epsilon_{ij} \qquad \mathrm{is\, not\, allowed\,in\,O_h},
\ee
where we explicitly write $\tau_0$ as the identity operator in the valley space. Third, the inversion symmetry also forbids strain-induced valley splitting,
\be
\delta H_\mathrm{low} \propto \tau_z \epsilon_{ij} \qquad \mathrm{is\, not\, allowed\,in\,O_h}.
\ee
However, one only needs to break the inversion symmetry to split the valleys by shear strain,
\be
\delta H_\mathrm{low} \propto (\tau_z \epsilon_{xy} +\mathrm{c.p.}) \qquad \mathrm{is\, allowed\,in\,T_d},
\ee
where c.p.~stands for cyclic permutation of the Cartesian indices. Using shear strain to induce valley splitting was proposed recently in Ref.~\cite{adelsberger_valley-free_2024}. One can consider other subgroups of $O_h$, such as $D_{2d}$ corresponding to the $X$ point of the Brillouin zone that was analyzed in Ref.~\cite{hensel_cyclotron_1965}. In this case, the three terms summed in the previous equation (including the term $\tau_z \epsilon_{xy}$) can appear individually, that is, with different prefactors.

\begin{figure}
	\centerline{\includegraphics[width=\linewidth]{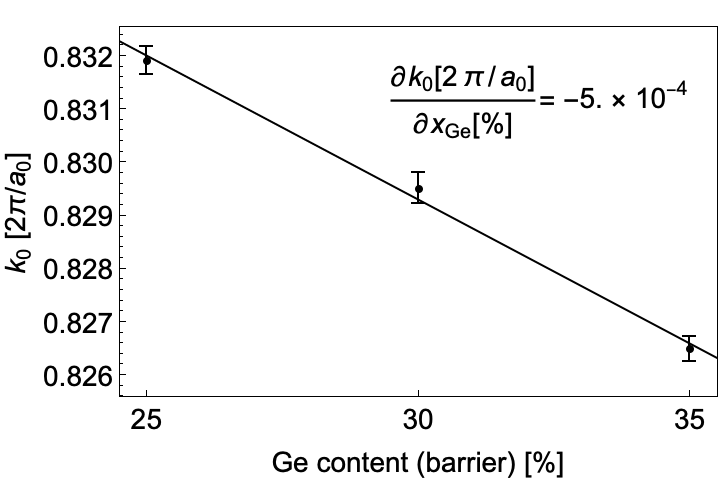}}
    \caption{
    \textbf{Change of the conduction band minimum with strain.} The points with error bars are values of $k_0$ taken from Tab.~\ref{tab:fittedParameters}. The line is a linear fit. We remind here that the constant $a_\mathrm{Si}$ refers to the monolayer distance in the strained silicon quantum well, rather than the value in the bulk Si crystal. This is a more convenient unit, as the natural unit for the monolayer positions given in Eq.~\eqref{eq:discreteGrid}. On the other hand, this definition means that $a_\mathrm{Si}$ changes with strain. We plot the equivalent Ge concentration in the barrier on the x axis, instead of the externally imposed strain actually used in our numerics.
    The slope fitted here on valley splittings in heterostructures is in good correspondence
with the shift of the conduction band minimum in the tight-binding band structures of bulk, 
biaxially strained silicon.
}
    \label{fig:shiftK0}
\end{figure}

After these checks, we investigate our case, which is a combination of $\mathbf{k}$, $\boldsymbol{\tau}$, and the strain tensor $\boldsymbol{\epsilon}$. We find that such terms are indeed allowed already in the group $O_h$ (and thus will be allowed for all its subgroups representing systems with lower symmetry). Specifically,
\be
\label{eq:allowedTerms}
\delta H_\mathrm{low} \propto 
\left. 
\begin{tabular}{c}
\vspace{0.1cm}
$\tau_z k_z \epsilon_{zz} +\mathrm{c.p.},$\\
\vspace{0.1cm}
$\mathrm{Tr}(\epsilon) \mathbf{k} \cdot \boldsymbol{\tau}$\\
$\mathbf{k} \cdot \boldsymbol{\epsilon} \cdot \boldsymbol{\tau}$
\end{tabular} 
\right\}
\qquad \mathrm{are\, allowed\,in\,O_h}.
\ee
While the entries in the second and third line are rather obvious invariants, the first line is somewhat nontrivial. In any case, all three lines contain the same answer to our question, showing that diagonal elements of the strain tensor induce a change in the conduction band minimum $k_0$. Combining the three terms, one can get
\be
\delta H_\mathrm{low} =A k_z \tau_z \epsilon_{zz} + B k_z \tau_z (\epsilon_{xx}+\epsilon_{yy}),
\ee
from which the change in the crystal momentum of the conduction band minimum under biaxial strain is
\be
\delta k_0 = -\frac{\epsilon_{zz}}{2k_0} \left( A-B \frac{c_{12}}{c_{11}}  \right),
\ee
where $c_{ij}$ are the stiffness constants. Although the values for the prefactors $A$ and $B$ can not be obtained from symmetry analysis, we conclude that there is no reason based on symmetry to expect that the value $k_0$ will not respond to strain.

\subsection{Results from numerics}

We now turn to numerics. We have repeated the analysis described in the main text for two values of externally imposed strain, $\pm 0.2\%$, expected to arise from changing the Ge concentration in the barriers by $\pm 5\%$. The results are given in Tab.~\ref{tab:fittedParameters} and plotted in Fig.~\ref{fig:shiftK0}. First, the change in $k_0$ is proportional to the change in strain. A linear relation is in line with the conclusion of symmetry analysis. Second, increasing the Ge concentration in the barrier by 10\% changes $k_0$ by -0.005. This change is small, and we conclude that tuning $k_0$ by strain most likely does not offer a decisive advantage for valley splitting unless the Ge barrier content can be varied extensively (say, by tens of percent).

We remind that the values of $k_0$ quoted above and given in Tab.~\ref{tab:fittedParameters} are expressed in units $2\pi/ a_\mathrm{Si}$, where the ``lattice constant'' $a_\mathrm{Si}$ is taken as the actual monolayer distance of the quantum well interior (mostly Si) along the growth direction. It also changes with strain. The change (with strain) of $k_0$ expressed in units of $1/\mathrm{nm}$ would be different.
However, for the phases entering the {\MEFFname}, the dimensionless $k_0$ is more convenient, which is why we keep it expressed in a unit that changes with strain. 

\section{Effect of atom count fluctuations}
\label{app:atomCountFluctuations}
In the code we use, both tight binding and DFT, the number of atoms of a given isotope can be controlled precisely. In the tight-binding simulations, in the Ge doping layer $n$ with nominal concentration $\rho_n$, we typically fix the number of Ge atoms to the integer closest to $\rho_n N_n$, where $N_n$ is the total atom count in a monolayer. Since $N_n$ is finite, being about 500 in our tight-binding simulations, if the Ge content were sampled randomly (each atom of the lattice replaced by Ge independently and randomly with probability $\rho_n$), the actual count of Ge atoms in a monolayer would fluctuate. We have shown in Ref.~\cite{Cvitkovich_VS_2} that the spread resulting from atom count fluctuations is about one order larger than from in-plane disorder alone (we have confirmed it also here by implementing the atom-count fluctuations in the data underlying Fig.~\ref{fig:real} and observing that it dominates the IQR spreads). If unknown, this fluctuating count would reduce the correlation between the observed valley splitting and the model prediction. We are interested in how big this effect is.

To this end, we use the model, Eq.~\eqref{eq:VSisM}, and, instead of keeping $\rho_n$ exactly at the nominal concentration, we sample the Ge atom count from a binomial distribution. We evaluate the correlation of the model prediction with the values from tight-binding data used in Fig.~\ref{fig:survey}. We find that for $N_n=500$ and $\rho_n=10\%$, the statistical fluctuations in the actual atom count in doping monolayers decrease the correlation coefficient between the model and numerics by additional 2\%.

\section{Comparison to a wiggle well}

\label{app:trouble}

We now revisit a comparison of the suggested patterns with the wiggle-well. So far, we have analyzed the discrete doping patterns based on valley splittings observed in tight-binding simulations and concluded that they appear promising compared to those observed in experiments with wiggle-well wafers. One can rightfully object that this is not a fair comparison. Ideally, one should compare the experimentally measured values for both. But since such data are not yet available, we take a shortcut and compare the values predicted by numerics. Also, since we find that the effective-mass model predictions correlate with tight-binding ones to 98\% (see Fig.~\ref{fig:fittingE0}), we will, for simplicity, in this section rely solely on the effective-mass model, expressed by Eq.~\eqref{eq:VSisM}.

\subsection{Common constraints on patterns}

\begin{figure}
	\includegraphics[width=0.8\linewidth]{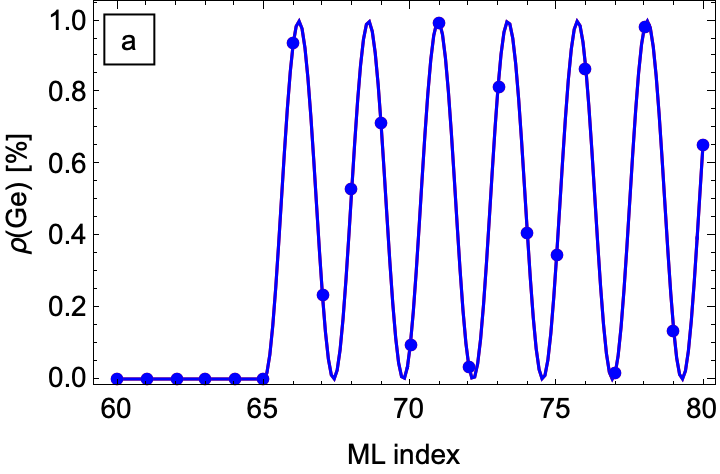}
	\includegraphics[width=0.8\linewidth]{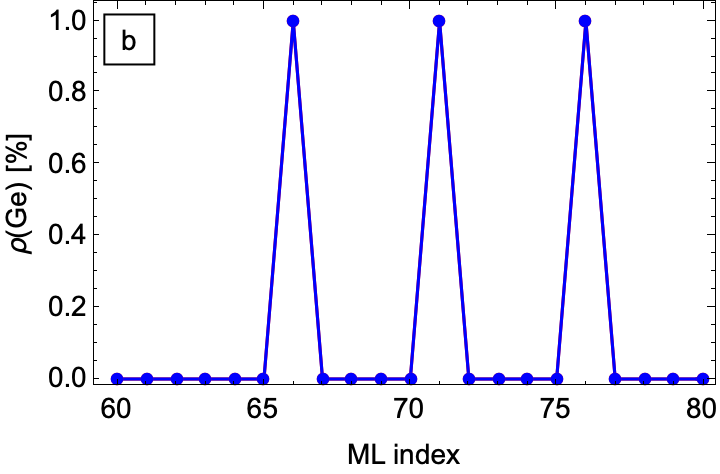}
         \includegraphics[width=0.8\linewidth]{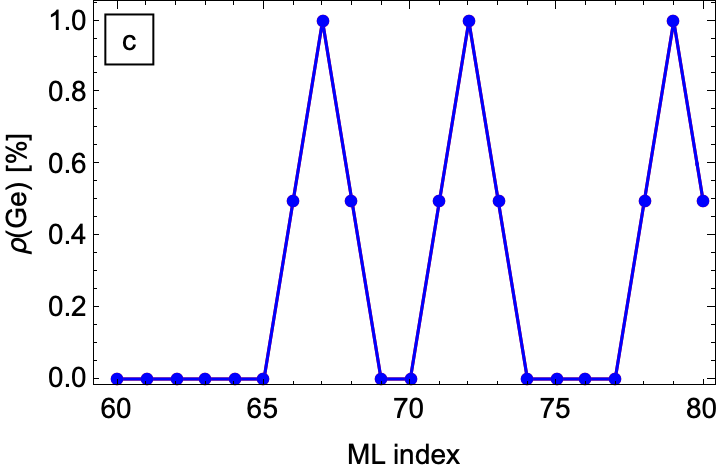}
	\includegraphics[width=0.8\linewidth]{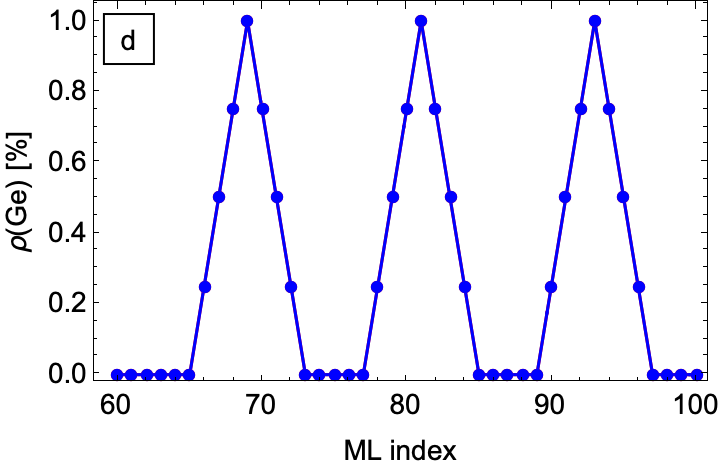}
    \caption{
    A few examples of doping profiles. They are constructed from continuous functions of position (curves), which set the average Ge concentration in a monolayer as the value of the continuous function evaluated on a discrete grid (dots). (a) The wiggle well. (b-d) Periodic patterns: (b) a peak with a width of 1 ML at every fifth monolayer, (c) a peak with a width of 3 ML at every fifth-then-seventh monolayer, (d) a peak with a width of 7 ML at every twelfth monolayer.
}
    \label{fig:patterns1}
\end{figure}

To compare the discrete patterns with the wiggle well expressed by Eq.~\eqref{eq:redHerring} in a meaningful way, one needs to consider some constraints as proxies of the limitations of the epitaxial growth that can actually be achieved in reality. We think it is natural to expect that, within certain limits specified below, if one can realize a doping pattern where the \textit{average} Ge doping concentration follows any jagged curve \textit{discretized on the finite grid representing monolayers}, then it should be possible to realize (again, a discretized version of) any smooth curve, including the sinusoidal one from Eq.~\eqref{eq:redHerring}.  Examples of such discretized curves are in Fig.~\ref{fig:patterns1}.

\begin{figure}
	\includegraphics[width=0.8\linewidth]{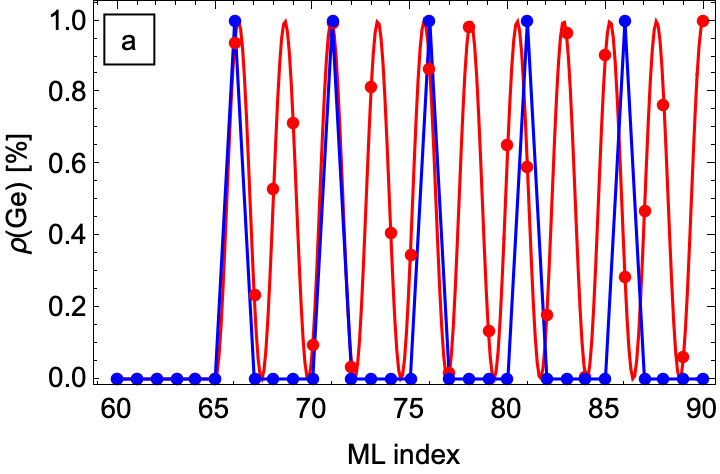}
	\includegraphics[width=0.8\linewidth]{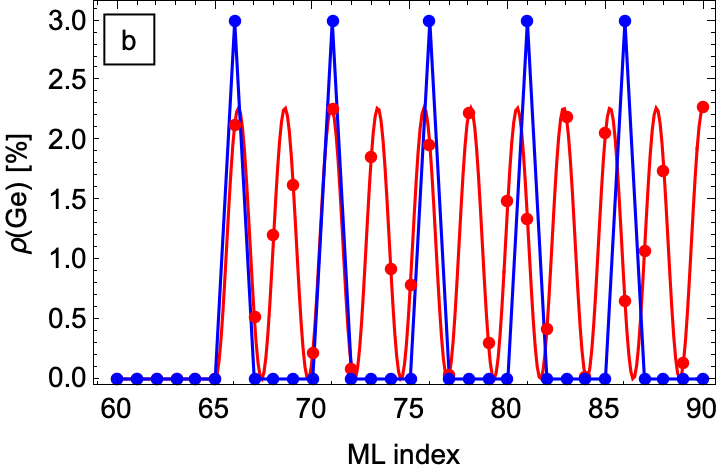}
         \includegraphics[width=0.8\linewidth]{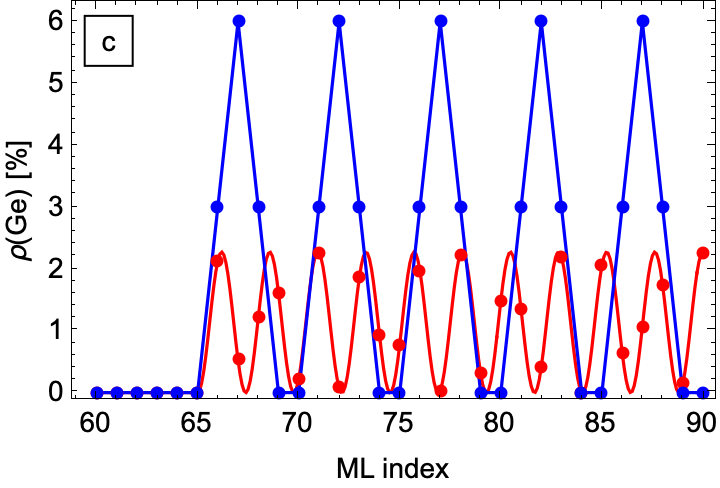}
	\includegraphics[width=0.8\linewidth]{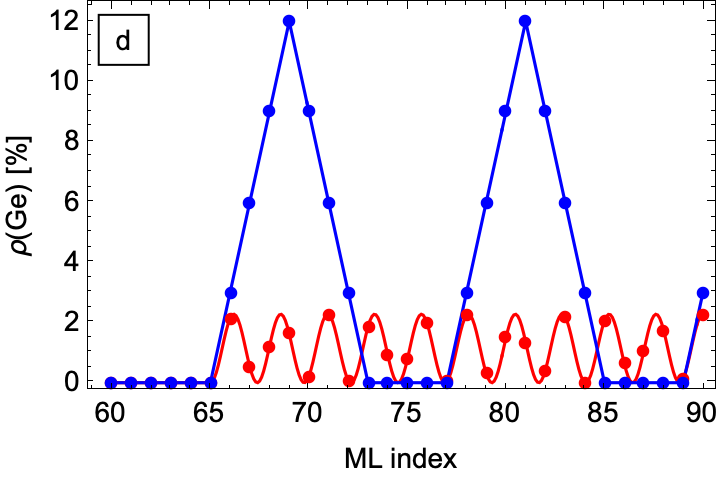}
    \caption{
    The wiggle well versus a few discrete doping profiles with various periods and peak widths. (a) With the constraint (made equal for both profiles) being the profile's maximum variation (between its minimum and maximum). (b-d) With the constraint being the profile's maximal gradient (concentration change between successive monolayers). We set the gradient to 3\% per monolayer, which is slightly less than the best reported value (of 4\% in Ref.~\cite{Gradwohl2024}.).}
    \label{fig:patterns2}
\end{figure}

We now examine constraints that allow a reasonable comparison of heterogeneous profiles. As an illustration, in Fig.~\ref{fig:patterns2}(a), we fix the maximal variation.\footnote{In a recent work, Ref.~\cite{thayil_optimization_2025}, the authors fix the total amount of Ge in the quantum well.} However, rather than the absolute value of Ge concentration, one expects that it is the change from one monolayer to the next that is limited during a quantum-well growth. Thus, limiting the achievable gradient seems a better constraint. Figures \ref{fig:patterns2}(b--d) show profile pairs with such a constraint. As expected, patterns with longer wavelengths can trade their lower density of doping elements (along the z axis) for a much larger maximal variation. This tradeoff will tend to compensate for the effectiveness of discrete patterns compared to the smooth wiggle well.

\subsection{Comparison of predicted valley splittings}

\begin{figure}
	\includegraphics[width=0.8\linewidth]{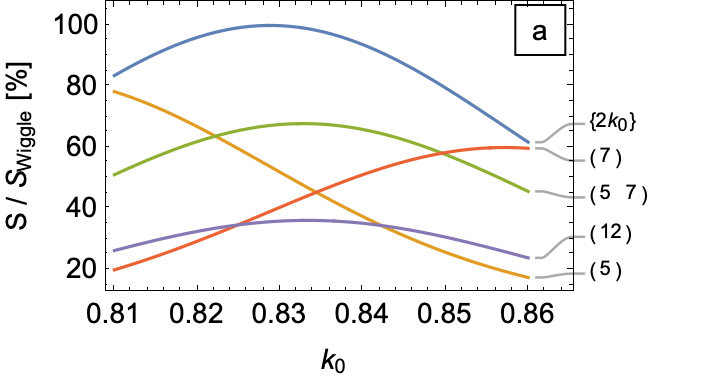}
	\includegraphics[width=0.8\linewidth]{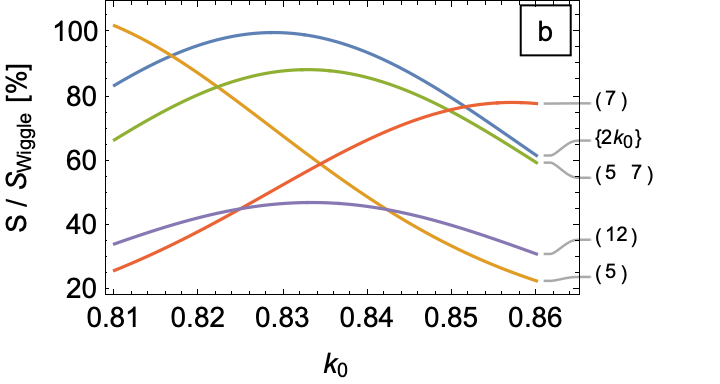}
         \includegraphics[width=0.8\linewidth]{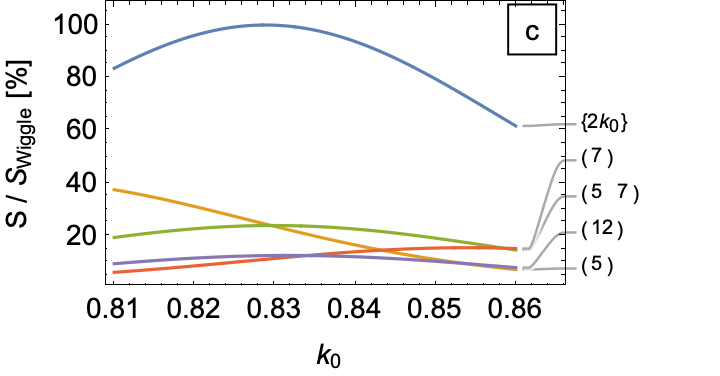}
	\includegraphics[width=0.8\linewidth]{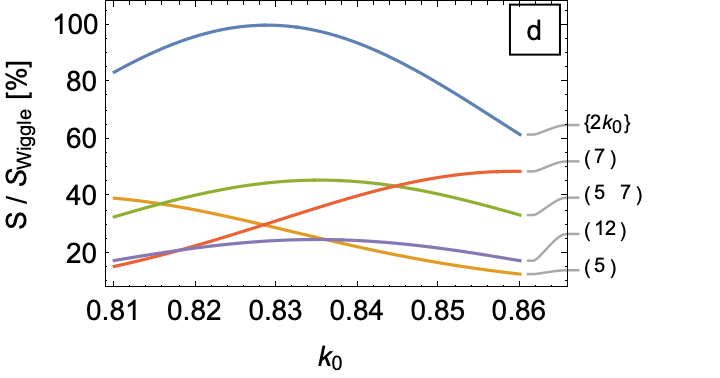}
    \caption{
Model predictions for various patterns. The included patterns are the wiggle well ("$2k_0$") and periodic ones with 5, 7, and 5-then-7 doping periods. In all panels, we assume the patterns are constructed assuming $k_0$ is 0.83, as a function of its actual value, which may be different (horizontal axis). We plot the ratio of the model-predicted structure factor (equivalently, valley splitting) to its value for a wiggle well assumed to be constructed for the actual value of $k_0$. (a) All patterns have the same maximal variation (the value is irrelevant for the ratios). (b-d) All patterns have the same maximal gradient. The discrete patterns have ML widths of one in (a) and (b), 2 in (c), and 3 in (d). 
}
    \label{fig:patterns3}
\end{figure}

Unfortunately, the compensation is not as high as one would hope for. The discrete patterns are less effective than the wiggle well unless they have a single-monolayer width. Figure \ref{fig:patterns3} summarizes this. Panel (a) is just for illustration, showing a comparison at a fixed maximal variation. Fig.~\ref{fig:patterns3}(b) shows that the discrete patterns are competitive when one fixes the pattern gradient, but they are not better than the wiggle well. Once the unrealistic assumption of their single-monolayer width is dropped, Fig.~\ref{fig:patterns3}(c)-(d), they become much worse than the wiggle well, by a factor that can be 3 or 5.  

Figure \ref{fig:patterns3}(c) shows that a halfwidth of two is actually worse than a halfwidth of three. One can understand this from the suppression factors defined in Eq.~\eqref{eq:suppression}. Let us consider a single doping peak with a width of three monolayers, and take $k_0=1$ for now. A single peak translates in Eq.~\eqref{eq:VSisM} into a vector $( -\frac{1}{2}, 1, -\frac{1}{2} )$ multiplying the density $w(z)$ on three consecutive lattice sites. One recognizes the discretized operator of the second derivative (times $-1/2$). Thus, multiplying by the phase factors and the doping weights effectively applies the second derivative. The structure factor becomes
\be
\label{eq:S as derivatives}
S \sim \sum_{p \in \mathrm{peaks}} d^2\partial_{zz}w(z_p),
\ee
where $d=a_\mathrm{Si}/4 = 0.13$ nm is the monolayer width.
If each point of the lattice were a doping peak, the sum would be zero exactly, no matter what the wave-function density profile $w(z)$. If the peaks are not dense in the lattice, one can estimate that the n-th derivative of the density scales with the inverse n-th power of the length $l_E= (\hbar^2/2m eE)^{1/3}$ associated with the electric field. For our parameters, $l_E \approx 2$ nm and the second-derivative terms in Eq.~\eqref{eq:S as derivatives} are of order $(d/l_E)^2$ and thus negligible. The finite value of the structure factor is dominated by the fact that the phase factor $e^{-2ik_0d}$ is not exactly -1. It is easy to see that for the discrete patterns that we consider, the discretized peaks turn the sum in Eq.~\eqref{eq:VSisM} into a sum analogous to Eq.~\eqref{eq:S as derivatives},  
\be
\label{eq:dzz}
S = \left| \sum_{p \in \mathrm{peaks}} e^{-2ik_0z_p}  \left( \sum_k c_k d^k\frac{\partial^k}{\partial_{z^k}} w(z) \right)_{z=z_p} \right|,
\ee
with several differential operators discretized on the lattice and $c_k$ being unspecified coefficients. Nevertheless, since the k-th derivative is suppressed by a factor $(d/l_E)^k$, they can all be dropped in the first approximation, leaving only the zero-th order term $c_0$, which we recognize as the factor $\alpha_{l,r}$ from Eq.~\eqref{eq:suppression}. Plugging in the triangular patterns from Fig.~\ref{fig:patterns1}, one gets that $\alpha_{2,2}$ is smaller than $\alpha_{3,3}$, which explains why the narrower patterns in Fig.~\ref{fig:patterns3}(c) are worse than the wider ones in Fig.~\ref{fig:patterns3}(d).

\subsection{How diffusion changes this picture}

In constructing the patterns in Fig.~\ref{fig:patterns1}, we considered the concentration gradient as the constraint bringing the candidate patterns on a common ground. That might seem too simplistic, so let us now consider diffusion. It is reasonable to assume that some form of diffusion occurs during the wafer growth or afterward across various device-fabrication steps. At first glance, the wiggle well patterns seem more vulnerable to diffusion than the discrete patterns, since they have shorter wavelengths. 

Let us then analyze the effects of the diffusion, expressed by  
\be
\label{eq:diffusionEquation1}
\partial_t \rho(z) = D \partial_{zz} \rho(z),
\ee
with $D$ the diffusion constant. The equation is simpler in the Fourier space, where the Fourier components of the concentration, $\rho(k)$, decouple
\be
\label{eq:diffusionEquation2}
\partial_t \rho(k) = - D k^2 \rho(k).
\ee
Assuming that the diffusion process acts for time $T$, the effects of the diffusion on the concentration, which we depict as $\rho \to \mathcal{D}[\rho]$, are given by the $k$-th Fourier component being suppressed by a factor
\be
\label{eq:diffusionFactor}
\rho(k) \to \mathcal{D}[\rho(k)] = \exp( -k^2 D T) \rho(k).
\ee
Indeed, features at shorter wavelengths are washed out exponentially faster than those at longer wavelengths.

However, the structure factor depends only on a single Fourier component: the one at $2k_0$. Concerning the \MEFFname, all patterns are suppressed by diffusion by the same factor. Namely, the operations of the diffusion $\mathcal{D}$ and the Fourier transform $\mathcal{F}$ commute, since they are both linear. The structure factor is proportional to 
\be
\mathcal{F}_{2k_0} [ \mathcal{D} [\rho] ] = \mathcal{D} [ \mathcal{F}_{2k_0} [ \rho] ] = \exp( -k_0^2 D T) \mathcal{F}_{2k_0} [ \rho ].
\ee
Diffusion imposes the same suppression factor on all patterns, irrespective of their shape. This conclusion applies to continuous space as well as discretized space: On a discretized lattice, the diffusion equation, Eq.~\eqref{eq:diffusionEquation1}, will be slightly different, leading to a slightly different form of the suppression factor in Eq.~\eqref{eq:diffusionFactor}. Nevertheless, this factor will be the same for all patterns, since $\mathcal{F}$ and $\mathcal{D}$ still commute, as long as the diffusion equation is diagonal in the Fourier space, Eq.~\eqref{eq:diffusionFactor}.

\subsection{Final remarks}

\begin{figure}
	\centerline{\includegraphics[width=0.8\linewidth]{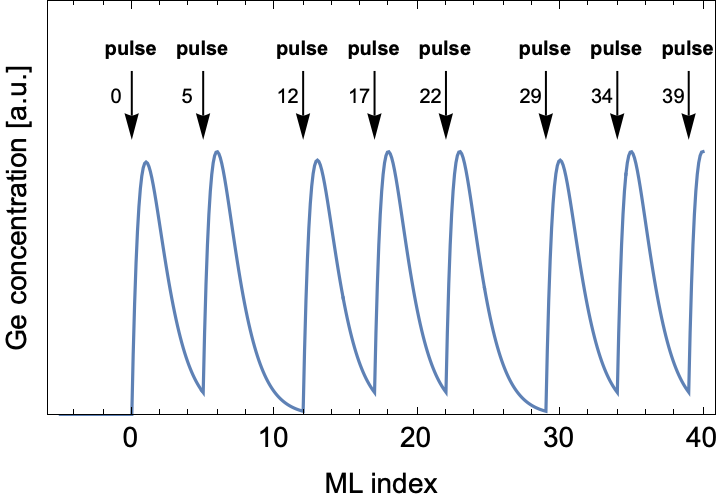}}
    \caption{Pulses of Ge doping synchronized with monolayers during the growth. Each time the monolayer thickness of the growing wafer, monitored in real time, reaches one of the prescribed values (here: 0, 5, 12, ... ), a Ge-doping pulse is injected into the chamber.}
    \label{fig:synchonizedPulses}
\end{figure}

The above comparison seems to cast a pessimistic picture concerning the discrete patterns as improvements over the wiggle well and its derivatives. Nevertheless, considering the immense complexity of the epitaxial growth process and the extent to which its details remain unknown \cite{pena_modeling_2024}, the above pattern comparison is likely inadequate and perhaps even naive. Instead of dwelling on such naive comparisons, we would like to point out one difference between any discrete pattern and a wiggle-well-like one that we deem crucial. Namely, all the discrete patterns we investigated might be seen as arising from a train of single impulses (injections of Ge dopants) during the epitaxial growth. It is not necessary that these impulses result in monolayer-thin doping. Rather, any doping pattern that \textit{relates to the discrete monolayer indexes} allows synchronization of injection timing with the material monolayer count, which can be monitored in real time during wafer growth. Figure \ref{fig:synchonizedPulses} expresses this idea in a schematic. Given the importance of the valley degeneracy problem for spin qubits, we believe this difference is worth testing experimentally. 

\section{Dependence on the Ge doping concentration}
\label{sec:nonlinearity}

Here we investigate how {\MEFFname} depends on the Ge doping concentration. One expects that our main result, Eq.~\eqref{eq:VSisM}, ceases to be valid once this concentration becomes high. In that regime, Ge doping changes the weights $w_m$ (an effect we neglected throughout when using Eq.~\eqref{eq:VSisM}) and the {\MEFFname} becomes a non-linear function of the Ge concentration. 

\begin{table}
 \begin{center}
\begin{tabular}{c@{\quad}c}
\toprule
Ge doping & $1 -$ correlation\\
concentration [\%] & coefficient [\%]\\
\midrule
10 & 1.96\\
5 & 0.46\\
1& 0.126\\
\bottomrule
\end{tabular}
\end{center}
\caption{\label{tab:nonlinearity}
\textbf{Model-data correlation versus doping concentration.} Correlation coefficient of the valley splitting observed in tight-binding data and the analytical formula, Eq.~\eqref{eq:VSisM}, over datasets similar to those investigated in the main text (about 700 random doping profiles). The Si well is doped by monolayer peaks with Ge concentration as given in the first column. The undoped quantum well is nominally identical throughout all sets and profiles, and the Ge concentration in the barrier is fixed to 30\%.
}
\end{table}

To this end, we have produced two additional datasets, analogous to the dataset investigated in the main text. All three datasets represent nominally equivalent clean (undoped) wells (with the Ge concentration in the barrier being 30\%) and differ by the Ge concentration in the doping peaks (being 10\%, 5\%, and 1\%, respectively). We repeat the fitting procedure described in App.~\ref{app:fittedParameters}, obtain data analogous to Fig.~\ref{fig:fittingE0} (not shown), and extract the correlation coefficient as an overall measure of how well the theory expressed in Eq.~\eqref{eq:VSisM} fits the tight-binding data. 

The results are in Tab.~\ref{tab:nonlinearity}. We observe a clear improvement in the theory-data match as the Ge doping decreases. At the lowest inspected doping of 1\%, the correlation coefficient reaches almost 99.9\%. This is a striking demonstration that the remaining small incorrelation is dominated by the non-linear effects, as we anticipated in Sec.~\ref{sec:real aspects}.

\begin{figure*}
	\centerline{
	\includegraphics[width=0.64\linewidth]{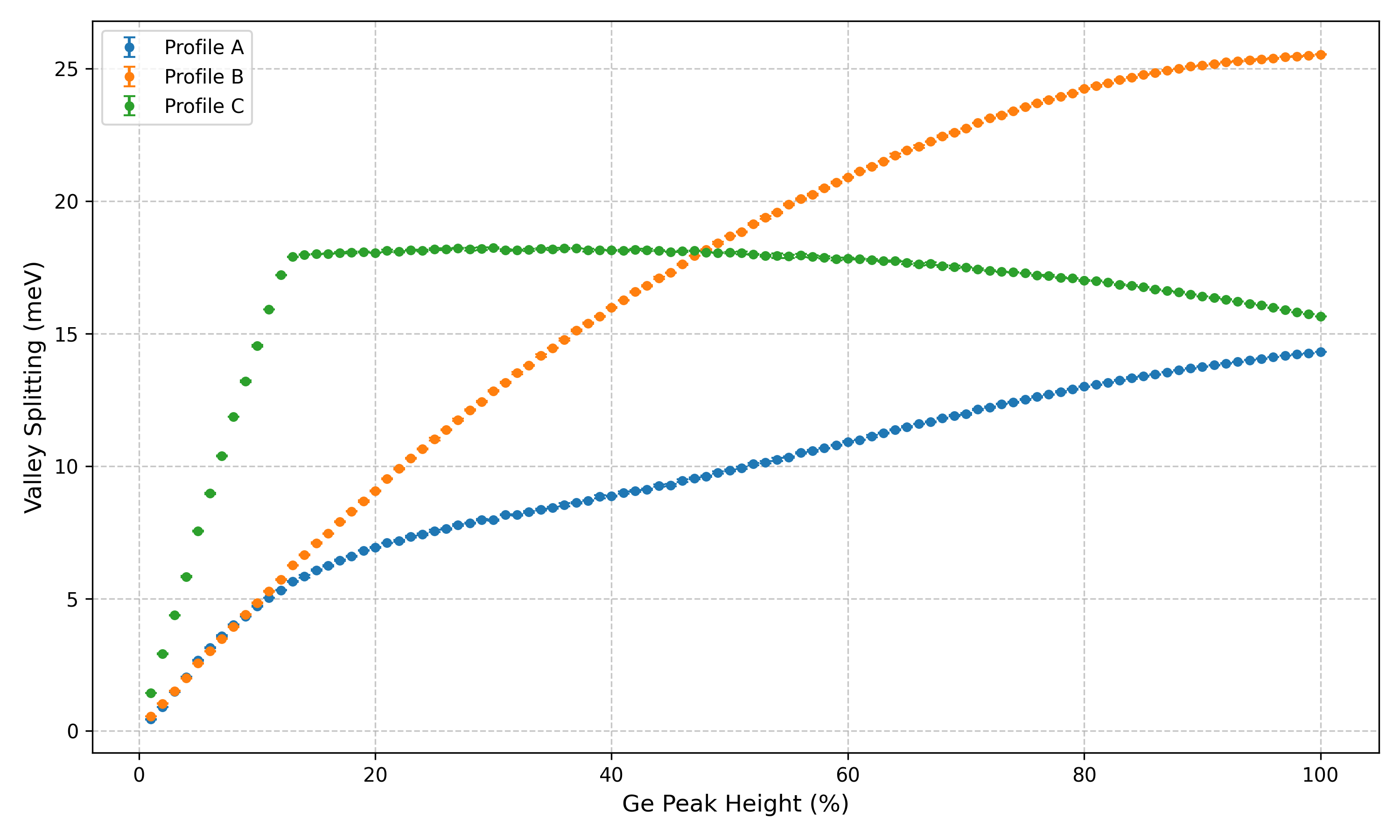}
	\includegraphics[width=0.34\linewidth]{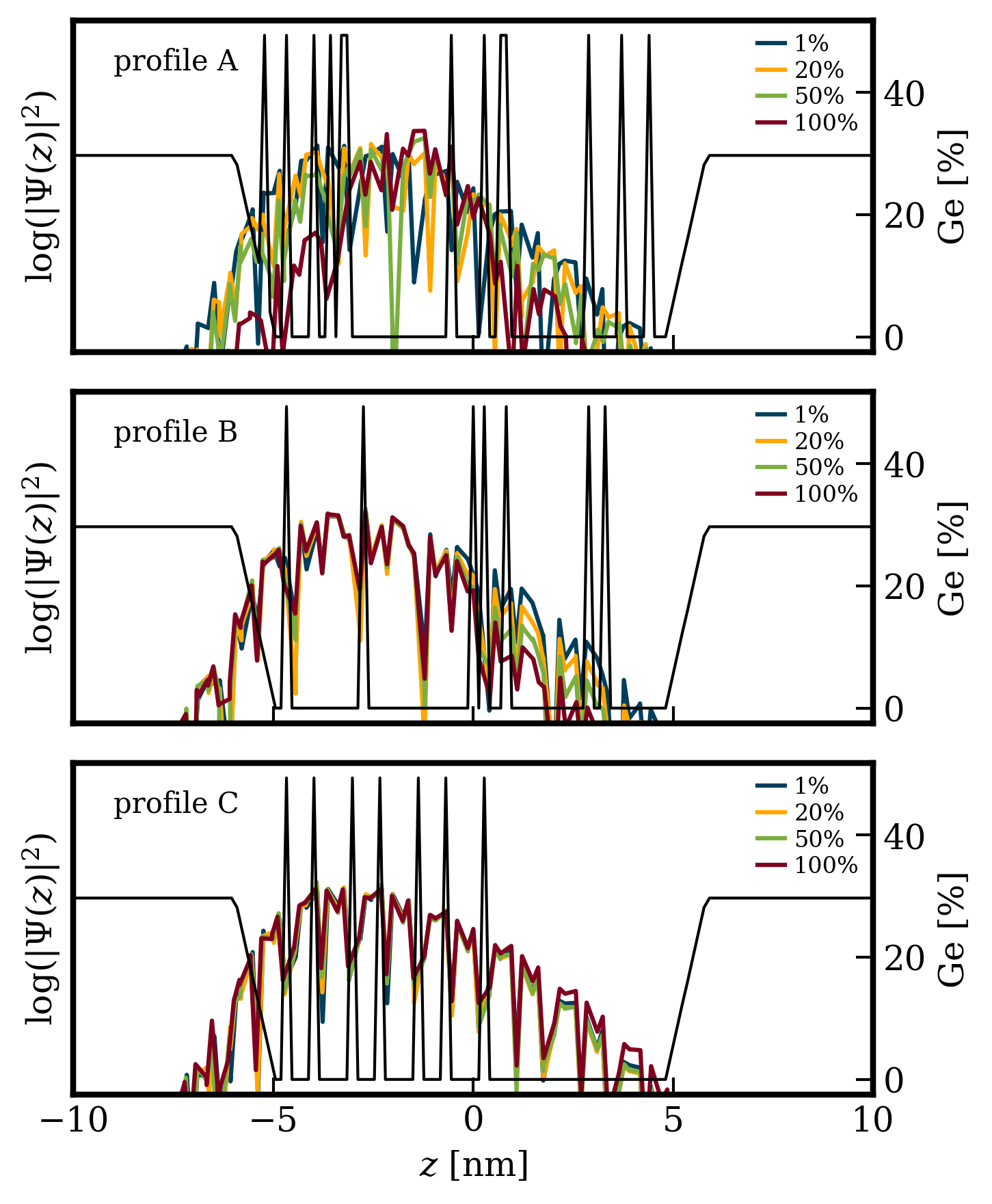}
	}
    \caption{Valley splitting versus Ge concentration for selected profiles. Left: Valley splitting (calculated by tight binding) for three doping profiles upon changing the Ge concentration of the doping monolayer peaks in the Si well (x axis). The Ge concentration in the barrier is fixed to 30\%. Right: The Ge profile defining the three doping profiles. Envelope wave functions (probability) are plotted for four doping concentrations, as denoted by the legends. The kink in the curve of profile $5$--$7$ marks an anticrossing with the lowest orbital state. The observed value matches the 17.8 meV excitation energy in a triangular potential for $E=5$ V/$\mu$m: $E_{j=2}-E_{i=1} = (2.04-1.17)\hbar^2/ m_l l_E^2$ with $l_E^3 = \hbar^2/2 m_l e E$.}
    \label{fig:threeProfiles}
\end{figure*}

To visualize these non-linear effects, we examine the valley splitting as a function of the doping concentration for three selected profiles: two random ones and the regular $5$--$7$ profile. The dependence is shown in Fig.~\ref{fig:threeProfiles}. The curves start approximately linearly at small concentrations and become sublinear at larger ones. This behavior is generic and has two sources. First, Ge-rich monolayers repel the electron wave function, so that the action of Ge atoms as scattering centers is partially offset by the decrease of the weights $w_n$ in Eq.~\eqref{eq:VSisM} that they cause. Second, as the valley splitting grows, the upper valley state approaches higher-lying (orbital) states and is repelled by them. In second-order perturbation theory, this shifts its energy by $|H_{ij}|^2/(E_i-E_j)$, where the coupling $H_{ij}$ depends on the doping profile, and so does the degree of the sublinearity. In the extreme case, there is a narrow anticrossing, as seen for the $5$--$7$ profile at a Ge concentration of about 13\%. Beyond this point, the second-lowest state is an orbital (here, vertical) excitation, and the plotted gap is no longer a valley splitting. We recall that we define the valley splitting as the energy gap between the two lowest states; its identification with a valley excitation, as well as the model in Eq.~\eqref{eq:VSisM}, are meaningful only as long as this gap stays well below the orbital excitation energy. (For the quantum well used in Fig.~\ref{fig:survey}, the lowest orbital excitation is at about 18 meV. In a realistic quantum dot, the lowest orbital excitation is an in-plane one, typically a few meV, which sets a practical ceiling on the useful valley splitting anyway.) If Ge concentrations at the doping peaks remain around or below 10\%, where the curves in Fig.~\ref{fig:threeProfiles} are close to linear, Eq.~\eqref{eq:VSisM} is an excellent approximation.

\newpage

\bibliography{my-clean_clean.bib,2025-Cvitkovich-ValleySplitting_clean.bib}

\end{document}